\documentclass[epj]{svjour} % Remove option referee for final version

\usepackage{graphicx} \graphicspath{ {./figures/}{./figures_py/} }
\usepackage{amsmath} % math mode
\usepackage{amssymb, bbm} % mathematica symbols
\usepackage{amstext} % text in math mode with \text{}
\usepackage{booktabs} \cmidrulekern=.3em % default .5em % nice tables
\usepackage{array}
\usepackage[utf8]{inputenc} %latin1

\renewcommand\Re{\operatorname{Re}}

\newcommand\numberthis{\addtocounter{equation}{1}\tag{\theequation}}
\newlength{\feynwidth} 
\newcommand{\figscale}{0.86}

\begin{document}

\title{Hyperons in nuclear matter from SU(3) chiral effective field theory}

\author{
S.~Petschauer\inst{1} \and
J.~Haidenbauer\inst{2} \and
N.~Kaiser\inst{1} \and
Ulf-G. Mei\ss{}ner \inst{2,3} \and
W.~Weise\inst{1,4}
}


\institute{
Physik Department, Technische Universit\"at M\"unchen, D-85747 Garching, Germany
\and
Institute for Advanced Simulation, J\"ulich Center for Hadron Physics, and\\
Institut f\"ur Kernphysik, Forschungszentrum J\"ulich, D-52425 J\"ulich, Germany
\and
Helmholtz-Institut f\"ur Strahlen- und Kernphysik and Bethe Center\\
for Theoretical Physics, Universit\"at Bonn, D-53115 Bonn, Germany
\and
ECT*, Villa Tambosi, 38123 Villazzano (Trento), Italy
}

\date{} %\date{Received: date / Revised version: date}

\abstract{
Brueckner theory is used to investigate the properties of hyperons in nuclear matter.
The hyperon-nucleon interaction is taken from chiral effective field theory at next-to-leading order with SU(3) symmetric low-energy constants. 
Furthermore, the underlying nucleon-nucleon interaction is also derived within chiral effective field theory.
We present the single-particle potentials of $\Lambda$ and $\Sigma$ hyperons in symmetric and asymmetric 
nuclear matter computed with the continuous choice for intermediate spectra.
The results are in good agreement with the empirical information. In particular, our calculation gives a 
repulsive $\Sigma$-nuclear potential and a weak $\Lambda$-nuclear spin-orbit force.
\PACS{
%        {PACS-key}{describing text of that key} %  \and
        {13.75.Ev}{} \and % {Hyperon-nucleon reactions}
        {14.20.Jn}{} \and % {Hyperons}
        {21.65.-f}{} % {Nuclear matter}
      }
}

\maketitle

\section{Introduction}
\label{sec:intro}

The interaction between hyperons (\(Y\)) and nucleons (\(N\)) is not only of interest by itself
but it constitutes also the input for microscopic calculations of few- and many-body systems
involving strange\-ness, such as hypernuclei or exotic neutron star matter.
Indeed, with regard to the latter, the observation of two-solar-mass neutron
stars \cite{Demorest2010,Antoniadis2013} provides strong restrictions for the
appearance of hyperons in neutron star matter or, more generally speaking, on the
in-medium properties of the hyperon-nucleon (\(YN\)) interaction \cite{Hell2014,Lonardoni2015,Vidana2015}. 
In particular, a sufficiently stiff equation-of-state (EoS) is required which does not leave much room
for the presence of hyperons in the dense cores of neutron stars.
A naive introduction of $\Lambda$-hyperons as an additional baryonic degree of
freedom would soften the EoS such that it fails
to support two-solar-mass neutron stars \cite{Djapo2010}.
There emerges a quest for repulsive $\Lambda$-nuclear forces at high baryon densities.
Purely phenomenological mechanisms have so far been invoked, e.g.\ through ad-hoc
vector meson exchange \cite{Weissenborn2012,Weissenborn2012a},
multi-Pomeron exchange \cite{Yamamoto2014}
or a suitably adjusted repulsive $\Lambda NN$ three-body interaction \cite{Lonardoni2015}.
Thus, there is an obvious need for a more systematic investigation of this issue.

Various phenomenological approaches based on meson-exchange models \cite{Holzenkamp1989,Reuber1994a,Rijken1998,Haidenbauer2005,Rijken2010} 
or quark models \cite{Kohno1999,Fujiwara2006} have been used to construct \(YN\) interactions in the past.
However, given the poor experimental data base, these interactions differ considerably from each other.
At the basic level the baryon-baryon interaction arises from the fundamental theory of the strong interaction, quantum chromodynamics (QCD).
Lattice QCD is approaching this problem via numerical simulations \cite{Aoki2012,Beane2011a,Beane2012}.
Alternatively, chiral effective field theory (chiral EFT) is exploiting the symmetries of QCD together with the appropriate low-energy degrees 
of freedom to construct the baryon-baryon interactions.
The \(YN\) interaction has been investigated at leading order (LO) \cite{Polinder2006} 
as well as next-to-leading order (NLO) \cite{Haidenbauer2013a,Petschauer2013a} within SU(3) chiral EFT. 
The \(YY\) interaction between all members of the baryon octet has also been studied \cite{Polinder2007,Haidenbauer2010a,Haidenbauer2015,Haidenbauer2016}.
The SU(2) chiral EFT framework, very successful in describing the low-energy nuclear forces to high precision \cite{Epelbaum2009,Machleidt2011},
has thus been extended to the strangeness sector.
Essential features of chiral EFT are that the results can be improved systematically by going to higher order in the power counting scheme, and that two- and three-baryon forces can be calculated in a consistent way. These properties make chiral EFT very suitable for describing baryonic forces.
The next-to-leading order \(YN\) potentials as derived from SU(3) chiral EFT include one- and two-pseudoscalar-meson (\(\pi,K,\eta\)) exchange diagrams and four-baryon contact terms with SU(3) symmetric low-energy constants (LECs).
Within this approach an excellent description of the experimental scattering data, comparable to the best phenomenological models, has been achieved \cite{Haidenbauer2013a}.

Numerous advanced techniques have been developed to investigate systems without and with strangeness using such microscopic interactions.
For instance in the few-body sector Faddeev-Yakubovsky theory \cite{Nogga2002,Nogga2014a} can provide very accurate results for systems with three or four particles.
Many-body approaches such as Quantum Monte Carlo calculations \cite{Gandolfi2007,Gandolfi2009,Lonardoni2013c}, or nuclear lattice computations \cite{Borasoy2007,Epelbaum2011,Meissner2014} work well in the nucleonic sector and can be extended to the strangeness sector.
Also many-body perturbation theory with chiral low-momentum interactions \cite{Holt2013,Coraggio2014a,Sammarruca2015} yields a good description of nuclear matter.
Concerning \(\Lambda\) and \(\Sigma\) hyperons in nuclear matter, specific long-range processes related to two-pion exchange between hyperons and nucleons in the nuclear medium have been studied in refs.~\cite{Kaiser:2004fe,Kaiser2005}.

Conventional Brueckner theory \cite{Brueckner1954,Brueckner1955,Day1967} at first order in the hole-line expansion, the so-called Bruecker-Hartree-Fock approximation, has been widely applied to calculations of
hypernuclear matter \cite{Rijken1998,Kohno1999,Schulze1998,Vidana1999} with the use of phenomenological two-body potentials.
The same approach is also employed in the context of neutron star matter \cite{Baldo1999,Schulze2006,Schulze2011}.
Recently, corresponding calculations of the properties of hyperons in nuclear matter have been also performed with a chiral \(YN\) interaction \cite{Haidenbauer2015a}. 
It has been demonstrated that the resulting in-medium properties of hyperons in symmetric nuclear matter around saturation density are consistent with the empirical information from hypernuclear phenomenology. 
In particular, a repulsive \(\Sigma\)-nuclear mean field and a very weak \(\Lambda\)-nuclear spin-orbit force have been found.

In the present work we extend the hypernuclear matter calculation of ref.~\cite{Haidenbauer2015a} in several ways:
(i) We employ an underlying nucleon-nucleon interaction that originates likewise from chiral EFT, 
(ii) we implement the continuous choice for intermediate-state spectra in Brueckner ladders, and (iii) we  
investigate isospin-asymmetric nuclear matter with variable neutron and proton densities.
Moreover, we compare our results to those obtained with the gap choice and with phenomenological \(YN\) potentials.
Our calculation of hyperons in infinite symmetric and asymmetric nuclear matter should be considered as starting point 
for future studies of hypernuclei and exotic neutron star matter within SU(3) chiral EFT.

Genuine three-baryon forces are disregarded in this work since the focus is on the hyperon-nucleon two-body interaction with special reference to its momentum dependence.
At NLO in the chiral power counting no three-baryon forces arise.
Chiral three-nucleon forces are, however, very important in order to get saturation of nuclear matter from chiral low-momentum two-body interactions treated in many-body perturbation theory \cite{Coraggio2014a}. The role of hyperon-nucleon-nucleon three-body forces for hypernuclei and exotic neutron star matter is a topic still 
under discussion, cf.\ refs.~\cite{Lonardoni2015,Vidana2010}.
This will be subject of future studies within the applied chiral effective field theory.
% A first step in the classification of the leading-order chiral three-baryon forces has been made in ref.~\cite{Petschauer2013b}.

The paper is organized as follows.
In sect.~\ref{sec:form} the construction of the baryon-baryon interaction from chiral effective field theory is reviewed and the employed non-relativistic Brueckner-Hartree-Fock method is described in some detail.
The results for hyperon single-particle potentials in isospin-symmetric and asymmetric nuclear matter, as well as pure neutron matter, are presented in sect.~\ref{sec:res}.
Section~\ref{sec:sum} gives a summary of our findings.

\section{Formalism}
\label{sec:form}

\subsection{Baryon-baryon interaction}

\begin{figure}[t]
\centering
\includegraphics[width=\feynwidth]{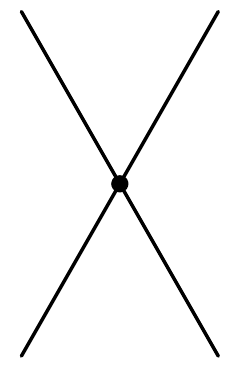} \quad
\includegraphics[width=\feynwidth]{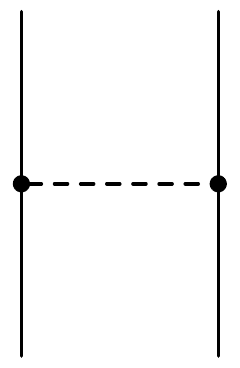}\\[.4\baselineskip]
\includegraphics[width=\feynwidth]{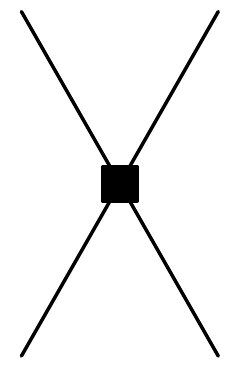} \quad
\includegraphics[width=\feynwidth]{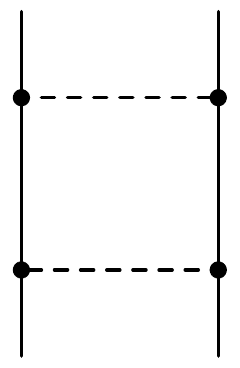}\quad
\includegraphics[width=\feynwidth]{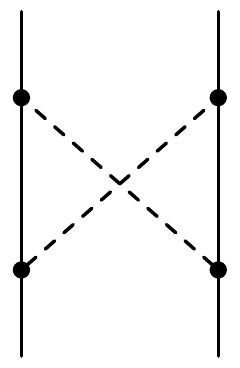}\quad
\includegraphics[width=\feynwidth]{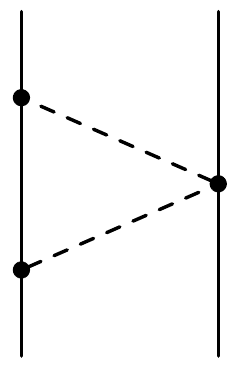}\quad
\includegraphics[width=\feynwidth]{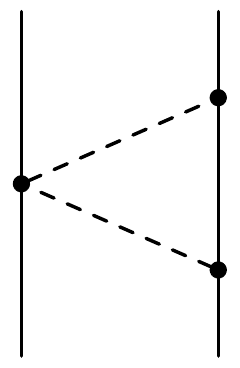}\quad
\includegraphics[width=\feynwidth]{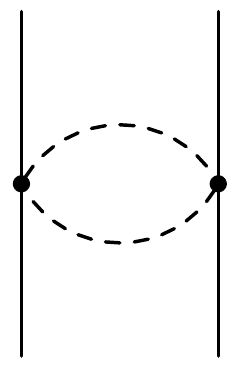}
\caption{Leading and next-to-leading order diagrams contributing to the baryon-baryon interaction potential. Solid and dashed lines denote octet baryons (\(N,\Lambda,\Sigma\)) and mesons (\(\pi,K,\eta\)), respectively.}
% A contact vertex proportional to \(q^2\) is symbolizes by a squared box.
\label{fig:pwrcounting}
\end{figure}

For the description of the hyperon-nucleon interaction we use SU(3) chiral effective field theory up to next-to-leading order with the Weinberg power counting applied to the potential, as reported in detail in \cite{Polinder2006,Haidenbauer2013a}.
At LO one-meson exchange diagrams and non-derivative four-baryon contact terms contribute to the potential. At NLO additional contact terms and two-meson exchange diagrams at the one-loop level arise, cf.\ fig.~\ref{fig:pwrcounting}.
The contact terms represent the unresolved short-distance dynamics, and the corresponding low-energy constants (LECs) are fitted to low-energy \(YN\) scattering data and the hypertriton (\({}^3_\Lambda \mathrm{H}\)) binding energy.
Given the sparse experimental information (in total 35 cross-section data points and a capture ratio) SU(3) flavor symmetric LECs have been used, while SU(3) symmetry breaking is incorporated through the physical masses of the exchanged pseudoscalar mesons (\(\pi,K,\eta\)).
Additional constraints on the \(P\)-wave potentials come from the nucleonic sector via SU(3) symmetry.

\begin{table}[t]
\caption{Contact terms and threshold parameters for the $^1S_0$ and
$^3S_1$-$^3D_1$ $NN$ partial waves for various cutoffs.
The values of the $\tilde C$'s are in
$10^4$ ${\rm GeV}^{-2}$, the ones of the $C$'s, in $10^4$ ${\rm GeV}^{-4}$;
the values of $\Lambda$ in MeV. The scattering length $a$ and the effective
range $r$ are in fm, the deuteron binding energy $E_d$ in MeV.
The empirical values are 
$a_{^1S_0}=-23.739$~fm, $r_{^1S_0}=2.68$~fm, 
$a_{^3S_1}=5.420$~fm, $r_{^3S_1}=1.753$~fm \cite{NijPWA}
and $E_d=-2.224575(9)$~MeV.
}
\label{tab:N1}
%\vspace{0.2cm}
\centering
\begin{tabular}{crrrr}
\toprule
{$\Lambda$} & $500$ & $550$& $600$& $650$ \\
\cmidrule(lr){1-1}\cmidrule(lr){2-5}
$\tilde C^{27}_{^1S_0}$   &$-0.1539$  &$-0.1017$ &$-0.0153$ &$0.1301$ \\
$C^{27}_{^1S_0}$          &$2.313$    &$2.326$   &$2.326$   &$2.328$ \\
%\addlinespace
$\tilde C^{10^*}_{^3S_1}$ &$-0.2100$  &$-0.1493$  &$0.0166$  &$0.2059$  \\
$C^{10^*}_{^3S_1}$        &$0.2977$  &$0.3139$ &$0.5109$  &$0.4899$  \\
$C^{10^*}_{^3S_1-\,^3D_1}$&$-0.2767$  &$-0.2896$ &$-0.2422$  &$-0.2234$ \\
\cmidrule(lr){1-1}\cmidrule(lr){2-5}
$a_{^1S_0}$                 &$-23.8$ &$-23.8$ &$-23.8$ &$-23.7$ \\
$r_{^1S_0}$                 &$2.81$ &$2.75$ &$2.68$ &$2.62$ \\
$a_{^3S_1}$                 &$5.42$ &$5.43$ &$5.42$ &$5.43$ \\
$r_{^3S_1}$                 &$1.81$ &$1.76$ &$1.72$ &$1.67$ \\
\cmidrule(lr){1-1}\cmidrule(lr){2-5}
$E_d$                  &$-2.257$ &$-2.213$ &$-2.193$ &$-2.145$ \\
\bottomrule
\end{tabular}
\end{table}

\begin{figure*}[t!]
\centering
\mbox{}
\includegraphics[width=.625\columnwidth]{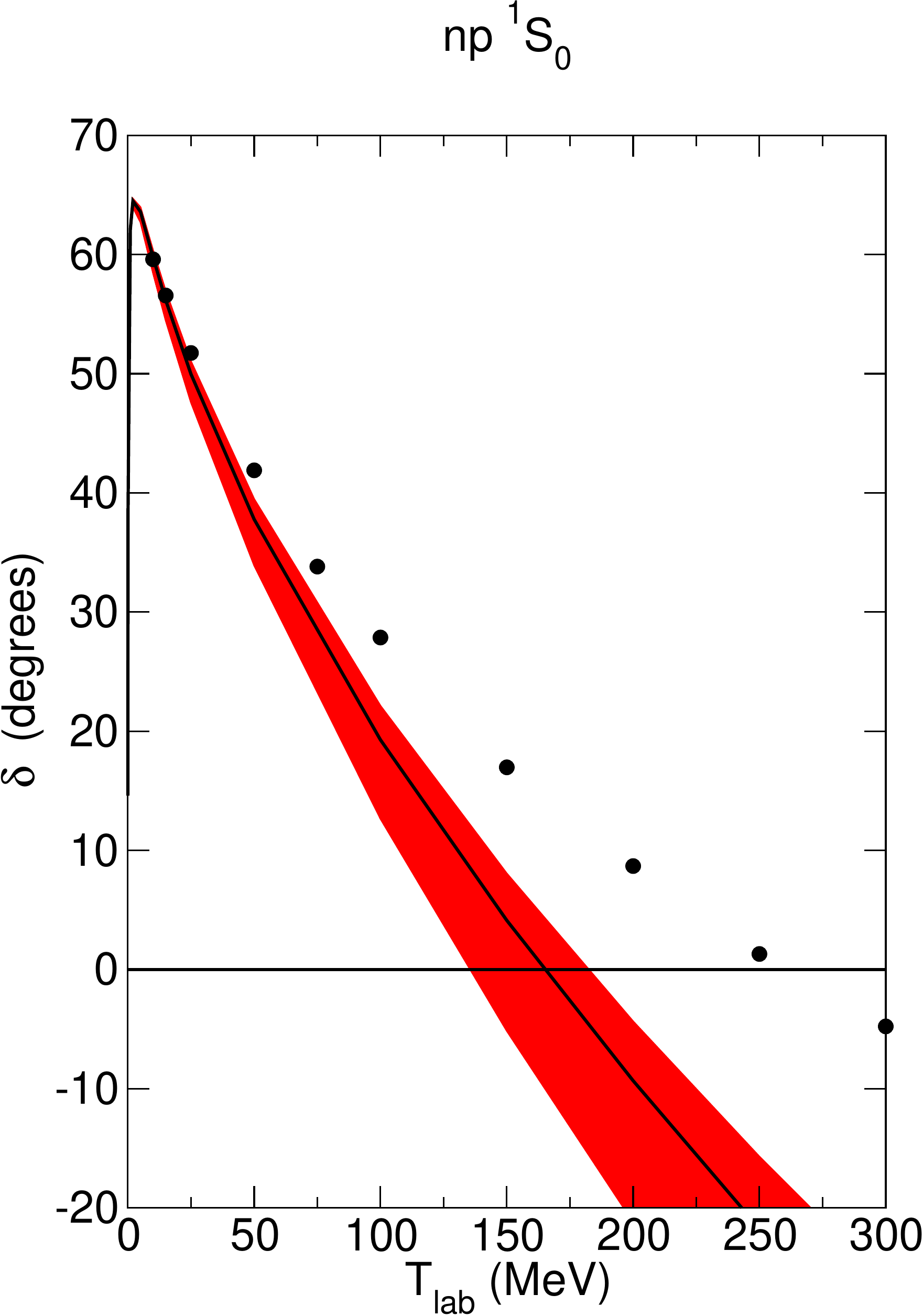} \qquad \qquad
\includegraphics[width=.625\columnwidth]{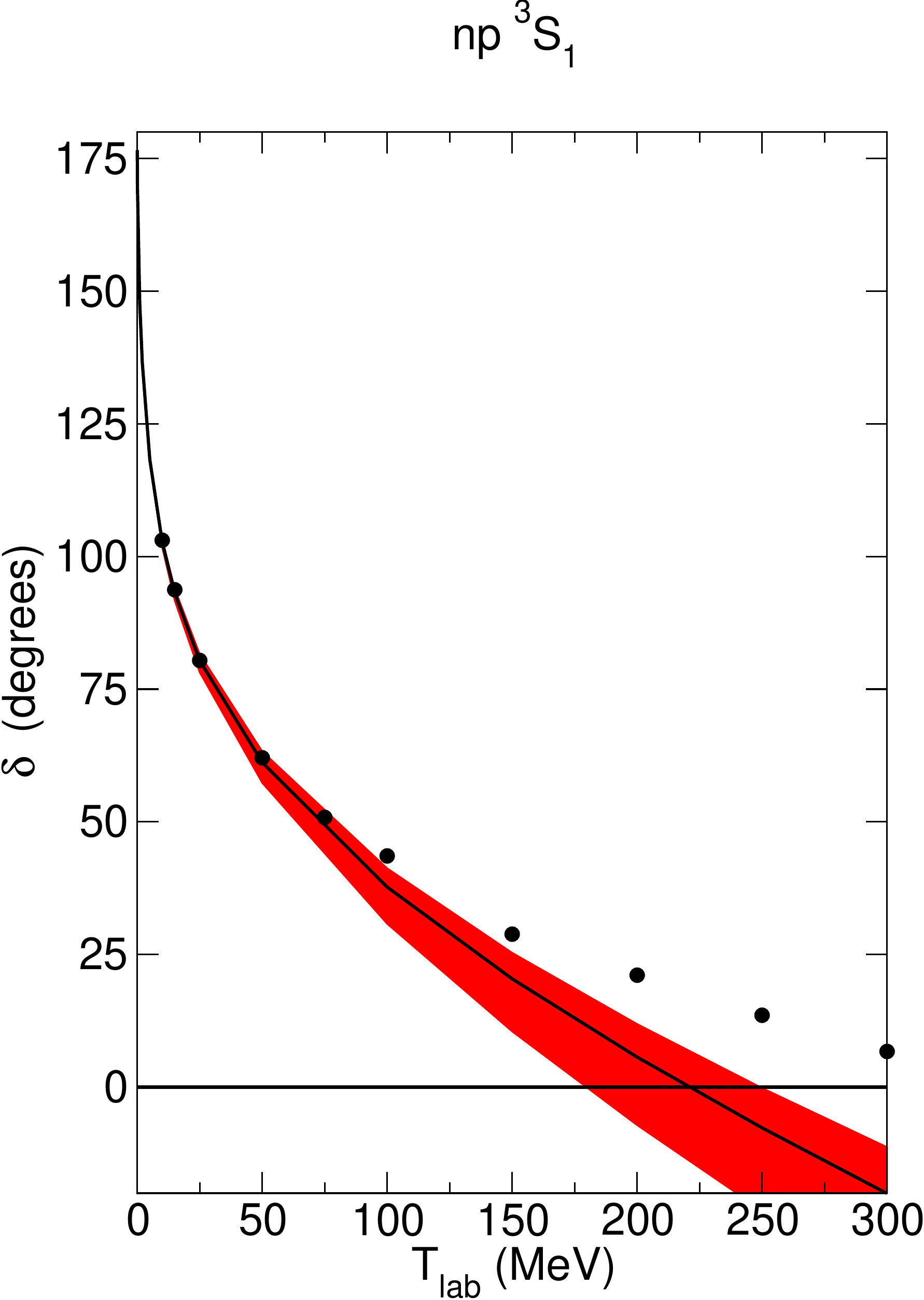}
\caption{Nucleon-nucleon phase shifts for the \({}^1S_0\) and \({}^3S_1\) partial wave. 
The band represents the variation of our NLO results with the cutoff, see text. 
The black line is obtained with a cutoff \(\Lambda = 600\text{ MeV}\). 
The circles denote the results from the GWU single-energy $np$ partial wave analysis \cite{SAID}. 
}
\label{fig:Nphase}
\end{figure*}

In order to obtain the reaction matrix (or \(T\)-matrix), the (two-particle irreducible) potentials are inserted into a regularized Lipp\-mann-Schwinger equation, which involves coupled partial waves as well as coupled two-baryon channels.
The coupled-channel Lippmann-Schwinger equation in the particle basis reads after partial-wave decomposition 
\begin{align*}
T^{{\rho''\rho'},J}_{{\nu''\nu'}}&(k'',k';\sqrt{s}) = V^{{\rho''\rho'},J}_{{\nu''\nu'}}(k'',k')\\
& +\sum_{{\rho},{\nu}}\int_0^\infty
\! \frac{\mathrm dk\,k^2}{(2\pi)^3}
\, V^{{\rho''\rho}\, ,J}_{{\nu''\nu}}(k'',k) \\
&\qquad\qquad\times \frac{2\mu_{\nu}}{k_{\nu}^2-k^2+i\epsilon}T^{{\rho\rho'},J}_{{\nu\nu'}}(k,k';\sqrt{s})\,,
\numberthis
\label{eq:LSE}
\end{align*}
where \(J\) denotes the conserved total angular momentum, \(\nu\) labels the two-particle channels (\(\Lambda p, \Sigma^+n,\Sigma^0p,\)\dots), and \(\rho\) represents the partial waves (\({}^1S_0, {}^3P_0,\dots\)).
Furthermore  \(\sqrt s\) is the center-of-mass energy and \(\mu_{\nu}\) stands for the pertinent reduced baryon mass.
The on-shell momentum \(k_\nu\) is determined by
\(\sqrt{s}=\sqrt{M^2_{B_{1,\nu}}+k_{\nu}^2}+\sqrt{M^2_{B_{2,\nu}}+k_{\nu}^2}\).
Note that high-energy components of the potential are cut off by a regulator function of the form \(f_R(\Lambda) = \exp[-(k'^4+k^4)/\Lambda^4]\), 
as done likewise in the nucleonic sector \cite{Epelbaum2004}.
In the following the cutoff \(\Lambda\) is varied in the range of \((500\linebreak[0]\dots\linebreak[0]650)\) MeV for NLO and \((550\dots700)\) MeV for LO,
i.e. in a range similar to that adopted for the nucleon-nucleon (\(NN\)) interaction in ref.~\cite{Epelbaum2004}. 
The resulting bands give an indication for the cutoff dependence and thus provide a lower bound on the theoretical uncertainty.
Recently, improved schemes to estimate the theoretical error were proposed and applied to the \(NN\) interaction \cite{Furnstahl2014,Epelbaum2015,Furnstahl2015}.
However, one has to go to higher orders in the chiral expansion if one wants to employ these schemes in a meaningful way. 
Therefore, we refrain from performing such an uncertainty quantification for the NLO interactions employed in the present study. 

A simultaneous description of the \(NN\) and \(YN\) interactions with SU(3) symmetric LECs is not possible at NLO \cite{Haidenbauer2013a} 
(due to the strong correlation between the $^1S_0$ partial wave in the \(NN\,(I=1)\) and \(\Sigma N\,(I=3/2)\) channels imposed by SU(3) symmetry).  
Therefore, we use different sets of LECs in the \(NN\) and \(YN\) sectors.
For the \(YN\) interaction the set given in refs.~\cite{Haidenbauer2013a} and \cite{Haidenbauer2015a} 
is used\footnote{Note that in order to be consistent with the definitions in eq.~(18) of ref.~\cite{Haidenbauer2015a} the constants \(C^{8_s8_a}\) 
in Table 1 have to be multiplied with a factor 2.}, 
where in the latter work the contact term that gives rise to an antisymmetric spin-orbit force in the \(YN\) interaction, 
responsible for spin singlet-triplet transitions, 
is already fitted to the weak \(\Lambda\)-nuclear spin-orbit interaction \cite{Gal:2010xn,Botta2012}.
The \(NN\) interaction is based on the same diagrams and contact terms as given in \cite{Haidenbauer2013a}, but with different LECs, compiled in table~\ref{tab:N1}.
Furthermore, table~\ref{tab:N1} provides the scattering length, the effective range and the deuteron binding energy, calculated from this chiral NLO \(NN\) interaction.
In fig.~\ref{fig:Nphase} the nucleon-nucleon phase shifts in the \({}^1S_0\) and \({}^3S_1\) partial waves are shown.
Note that the results are comparable to those (at NLO) in ref.~\cite{Epelbaum1999a} where SU(2) chiral EFT was used 
(cf. fig.~4 in that reference). 
At low energies they are in agreement with the empirical data.
However, at higher energies the results of the NLO interaction become too repulsive.

\subsection{First order conventional Brueckner theory}

In order to investigate the properties  of hyperons in nuclear matter we employ the conventional Brueckner theory at first order in the hole-line expansion \cite{Brueckner1954,Brueckner1955,Day1967}, the so-called Brueckner-Hartree-Fock approximation.
We focus on the single-particle potentials, i.e.\ the mean fields of hyperons in nuclear matter.
Our calculations are done in the particle basis and in the following we summarize the relevant formalism. 
For a more detailed introduction we refer the reader to ref.~\cite{Kohno1999} and also to refs.~\cite{Reuber1994a,Rijken1998,Schulze1998,Vidana1999}.

A central object of Brueckner theory is the  \(G\)-matrix representing the effective interaction between two particles in the medium.
The Brueckner reaction matrix or \(G\)-matrix is determined by solving the Bethe-Goldstone equation
\begin{equation}
G(\omega) = V + V \frac Q{e(\omega)+\mathrm i\epsilon} G(\omega)\,,
\end{equation}
with \(e(\omega)\) the energy denominator depending on the starting energy \(\omega\).
The Pauli blocking operator \(Q\) ensures that particles in intermediate states are outside their own Fermi sea.
The same potential \(V\) as in the Lippmann-Schwinger equation \eqref{eq:LSE} for free scattering is used.
Additional medium modifications of the two-meson exchange potential are not included.
These represent density dependent two-body forces arising from three-body interactions.
The medium effects come, therefore, solely from the Pauli operator \(Q\) in the Bethe-Goldstone equation and the density-dependent single-particle potential in the energy denominator \(e(\omega)\).
% In the nucleonic sector such additional medium modifications amount to only small corrections \cite{Saviankou2009}.

After angle-averaging, the Bethe-Goldstone equation decomposes into partial waves and reads for conserved values of the total angular momentum \(J\), total momentum \(\vec K\) and starting energy \(\omega\):
\begin{align*}
&G^{{\rho''\rho'},J}_{{\nu''\nu'}}(k'',k';K,\omega) =
V^{{\rho''\rho'},J}_{{\nu''\nu'}}(k'',k')\\
&\qquad+\sum_{{\rho},{\nu}}\int_0^\infty\!\! \frac{\mathrm dk\, k^2}{(2\pi)^3} \, V^{{\rho''\rho}\, ,J}_{{\nu''\nu}}(k'',k) \\
&\qquad\qquad\qquad\times 
\frac{\bar Q_{\nu}(K,k)}{\bar e_{\nu}(K,k;\omega)+i\epsilon}G^{{\rho\rho'},J}_{{\nu\nu'}}(k,k';K,\omega) \,.
\numberthis
\label{eq:BGE}
\end{align*}
As in eq.~\eqref{eq:LSE} the symbol \(\rho\) stands for the partial waves, \(\rho=(SL)\).
The (coupled) two-particle channels are \(\nu=(B_1B_2)\), with the baryons from the set: \(B_i \in \{n,\linebreak[0] p,\linebreak[0] \Lambda,\linebreak[0] \Sigma^+,\linebreak[0] \Sigma^0,\linebreak[0] \Sigma^-\}\). %,\Xi^0,\Xi^- % % \linebreak % \allowbreak
In the initial state the baryon \(B_2\) is within its own Fermi sea.
We introduce the total and relative momenta of two baryons \(B_1\) and \(B_2\) by
\begin{equation}
\vec K = \vec k_1 + \vec k_2\,,\quad
\vec k = \frac{\xi_{12}\vec k_1-\vec k_2}{1+\xi_{12}}\,,\quad
\xi_{12} = \frac{M_{2}}{M_{1}}\,.
\end{equation}

In eq.~\eqref{eq:BGE} we have applied the standard approximation replacing \(Q/e\) by the ratio of its angle-averages \(\bar Q/\bar e\).
The Pauli blocking operator involving the Fermi momenta \(k_F^{(1,2)}\) of the two baryon species is given by 
\begin{align*}
\bar Q_{\nu}(K,k) & =
\frac12\int_{-1}^1\!\mathrm d\cos\theta\
\Theta\left(|\vec k_1| - k_F^{(1)}\right)\Theta\left(|\vec k_2| - k_F^{(2)}\right) \\
 & = [0|\frac{[-1|z_1|1]+[-1|z_2|1]}2|1] \,,
\numberthis
\end{align*}
when averaged over the angle \(\theta\) between \(\vec K\) and \(\vec k\), 
with the shorthand notation \([a|b|c]\equiv\operatorname{max}(a,\operatorname{min}(b,c))\) introduced in \cite{Schulze1998}, and the arguments
\begin{align*}
z_1 & = \frac{1+\xi_{12}}{2kK}\left\{\left(\frac1{1+\xi_{12}}K\right)^2+k^2-(k_F^{(1)})^2\right\}\,, \\
z_2 & = \frac{1+1/\xi_{12}}{2kK} \left\{\left(\frac{\xi_{12}}{1+\xi_{12}}K\right)^2 + k^2-(k_F^{(2)})^2\right\} \,.
\numberthis
\end{align*}

The angle-averaged energy denominator is given by
\begin{align*}
\bar e_{\nu}(K,k;\omega) = \omega & - \frac{K^2}{2M_\nu}-\frac{k^2}{2\mu_\nu} -M_\nu \\
& - \Re U_{B_1}(\bar k_1)- \Re U_{B_2}(\bar k_2) \,,
\numberthis
\end{align*}
with total and reduced masses, \(M_{\nu}=M_1+M_2\) and \(\mu_{\nu}=M_1 M_2/(M_1+M_2)\). The angle-average is done for the arguments of the single particle potentials \(U_{B_i}\) of the intermediate baryons:
\begin{align*}
\bar k_1
&= \left(\tfrac1{(1+\xi_{12})^2} K^2 + k^2 + 2\tfrac1{1+\xi_{12}} K k\, \overline{\cos\theta}\right)^{1/2} \,, \\
\bar k_2
&= \left(\tfrac{\xi^2_{12}}{(1+\xi_{12})^2} K^2 + k^2 - 2\tfrac{\xi_{12}}{1+\xi_{12}} K k\, \overline{\cos\theta}\right)^{1/2} \,,
\numberthis
\end{align*}
with the mean directional cosine
\begin{align*}
\overline{\cos\theta} &
= \frac{ \int_{-1}^1\!\mathrm d\cos\theta\, \cos\theta\, Q(\vec K,\vec k) }{ \int_{-1}^1\!\mathrm d\cos\theta\, Q(\vec K,\vec k) } \\
& = \frac12\big([-1|z_2|1]-[-1|z_1|1]\big) \, ,
\numberthis
\end{align*}
where \(Q(\vec K,\vec k)\) is the exact Pauli blocking operator.
If two nucleons are involved, the previous expression for \(\overline{\cos\theta}\) would vanish (in symmetric nuclear matter), because of the almost equal masses, \(z_1\approx z_2\equiv z_0\).
Then the alternative angular average
\begin{align*}
\overline{\cos\theta} &= \sqrt{\frac{ \int_{-1}^1\!\mathrm d\cos\theta\, \cos^2\theta\, Q(\vec K,\vec k) }{ \int_{-1}^1\!\mathrm d\cos\theta\, Q(\vec K,\vec k) }}
= \frac1{\sqrt3}[0|z_0|1]
\numberthis
\end{align*}
is often used.

It is common practice to introduce a further simplification.
The squared momenta \(K^2=K^2(\vec k_1, \vec k)\) and \(k^2_2=k^2_2(\vec k_1, \vec k)\) entering the Bethe-Goldstone equation are replaced by their angle averages:
\begin{align*}
\bar{K}^2(k_1,k) & = \frac{\int_{|\vec k_2|\leq k_F^{(2)}}\!\mathrm d\cos\vartheta\, K^2(k_1,k,\cos\vartheta)}{\int_{|\vec k_2|\leq k_F^{(2)}}\!\mathrm d\cos\vartheta} \\
&=
(1+\xi_{12})^2\left[k_1^2+k^2-k_1k(1+[-1|x_0|1])\right] \,, \\
\bar k_2^2(k_1,k)
&=
\frac{\xi_{12}}{1+\xi_{12}}\bar K^2(k_1,k) + (1+\xi_{12})k^2 -\xi_{12}k_1^2 \,,
\numberthis
\end{align*}
where \(\vartheta\) is the angle between \(\vec k_1\) and \(\vec k\), and \(x_0\) means:
\begin{equation*}
x_0 = \frac{\xi_{12}^2k_1^2+(1+\xi_{12})^2k^2-(k_F^{(2)})^2}{2\xi_{12}(1+\xi_{12})k_1k} \,.
\numberthis
\end{equation*}
Note again, that the baryon \(B_2\) in the initial state is within its Fermi sea.

Finally, the single-particle potential of a baryon \(B_1\) due to the Fermi sea of the species \(B_2\) is calculated in the Brueckner-Hartree-Fock approximation as follows
% averaged of the spin projection \(\sigma_1\) of \(B_1\)
\begin{align*}
U_{B_1}^{(B_2)}(k_{1}) = &
\left(1+\delta_{B_1B_2}(-1)^{L+S}\right)
\frac{(1+\xi_{12})^3}{2} \\
& \times \sum_{J, \rho}(2J+1)
\int_{k_\mathrm{min}}^{k_\mathrm{max}}\! \frac{\mathrm dk\, k^2}{(2\pi)^3} \\
& \times
W(k_{1},k) \, G_{(B_1B_2)(B_1B_2)}^{\rho\rho,J}(k,k;\bar K,\omega_\mathrm{o.s.}) \,.
\numberthis
\label{eq:U}
\end{align*}
The full single-particle potential of a baryon \(B_1\) (\(B_1=n,p,\Lambda,\Sigma^{0,\pm}\)) is then given by the sum of the contributions from all baryons \(B_2\) (\(B_2=n,p\)) in the nuclear Fermi sea.
The weight function \(W(k_1,k)\) has the form
\begin{align*}
W(k_1,k) &= \frac1{4\pi}\int_{|\vec k_2|\leq k_F^{(2)}}\!\mathrm d\Omega_k = \frac12(1-[-1|x_0|1]) \,.
\numberthis
\end{align*}
The lower and upper integration boundaries of the relative momentum, \(k_\mathrm{min}\) and \(k_\mathrm{max}\), are determined from \(W(k_1,k) = 0\), which leads to
\begin{align*}
k_\mathrm{min} &= \operatorname{max}\left( 0, \frac{-k_F^{(2)}+\xi_{12} k_{1}}{1+\xi_{12}} \right) \,,\
k_\mathrm{max} = \frac{k_F^{(2)}+\xi_{12} k_{1}}{1+\xi_{12}} \,.
\numberthis
\end{align*}
The G-matrix elements in eq.~\eqref{eq:U} are calculated at the on-shell starting energy
\begin{align*}
\omega_\text{o.s.} &= E_{B_1}(k_{1})+E_{B_2}(\bar k_{2}) \\
E_{B_i}(k_i) &= M_i + \frac{k_i^2}{2M_i} + \Re U_{B_i}(k_i) \,.
\numberthis
\end{align*}
This makes the determination of the single-particle potentials dependent on the single-particle potential itself, 
and, therefore, eqs.~\eqref{eq:BGE} and \eqref{eq:U} have to be solved self-consistently.

In the so-called gap choice the single-particle potential is given by eq.~\eqref{eq:U} for \(k_1\leq k_F^{(1)}\) and set to zero for \(k_1>k_F^{(1)}\). 
Therefore, only the free particle energies of the intermediate states appear in the energy denominator of the Bethe-Goldstone equation since the Pauli blocking operator is zero for momenta below the Fermi momentum.
In the so-called continuous choice eq.~\eqref{eq:U} is used for the whole momentum range, hence the single-particle potentials enter also in the energy denominator.
The result of the Brueckner theory to all orders in the hole-line expansion should be independent of the choice for the intermediate-state spectrum, 
however, at leading order the results differ somewhat.
In the nucleonic sector at first order the continuous choice lies actually closer to the second order result of Brueckner theory, according to ref.~\cite{Song1998a}.
Moreover, the continuous choice for intermediate spectra allows for a reliable determination of the single-particle potentials including their imaginary parts \cite{Schulze1998}.
Unless stated differently, we use in this work the continuous choice.

\section{Results}
\label{sec:res}

In this section we present our results for the in-medium properties of hyperons, based on a $YN$ interaction derived 
from chiral EFT~\cite{Haidenbauer2013a}.
Additionally, for the ease of comparison, the \(G\)-matrix results obtained with two phenomenological $YN$ potentials, namely of
the J\"ulich~'04 \cite{Haidenbauer2005} and the Nijmegen NSC97f \cite{Rijken1998} meson-exchange models, are given.
Note that, like the EFT potentials, those \(YN\) interactions produce a bound hypertriton \cite{Nogga2014a}.

As mentioned before, the EFT \(NN\) and \(YN\) interactions involve different sets of low-energy constants.
For calculations with the LO and NLO hyperon-nucleon interaction we employ as the underlying nucleon-nucleon interaction 
the NLO version with the same cutoff.
In the case of the phenomenological \(YN\) interactions (J\"ulich~'04 and Nijmegen NSC97f) we use for the purpose 
of comparison the NLO chiral \(NN\) potential with a cutoff of 600 MeV.
In all calculations sums over partial waves up to \(J\)=5 are performed.

\begin{figure*}[htpb]
\centering
\includegraphics[scale=\figscale]{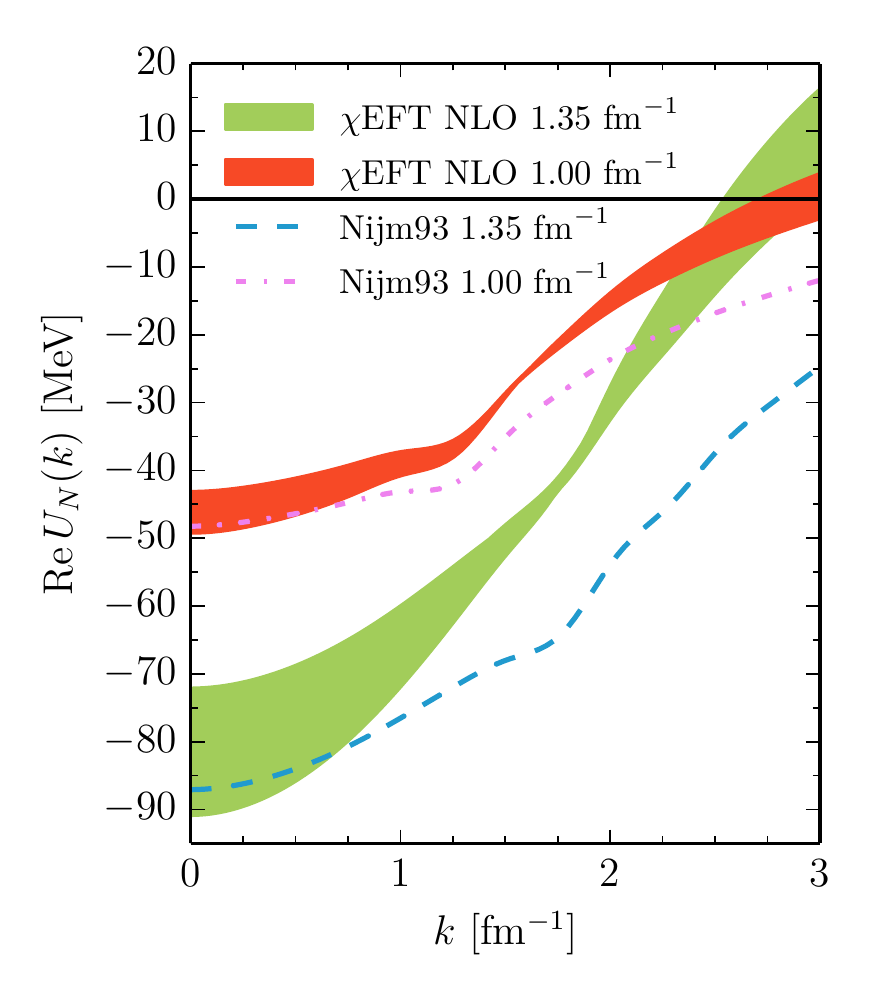} \qquad \qquad
\includegraphics[scale=\figscale]{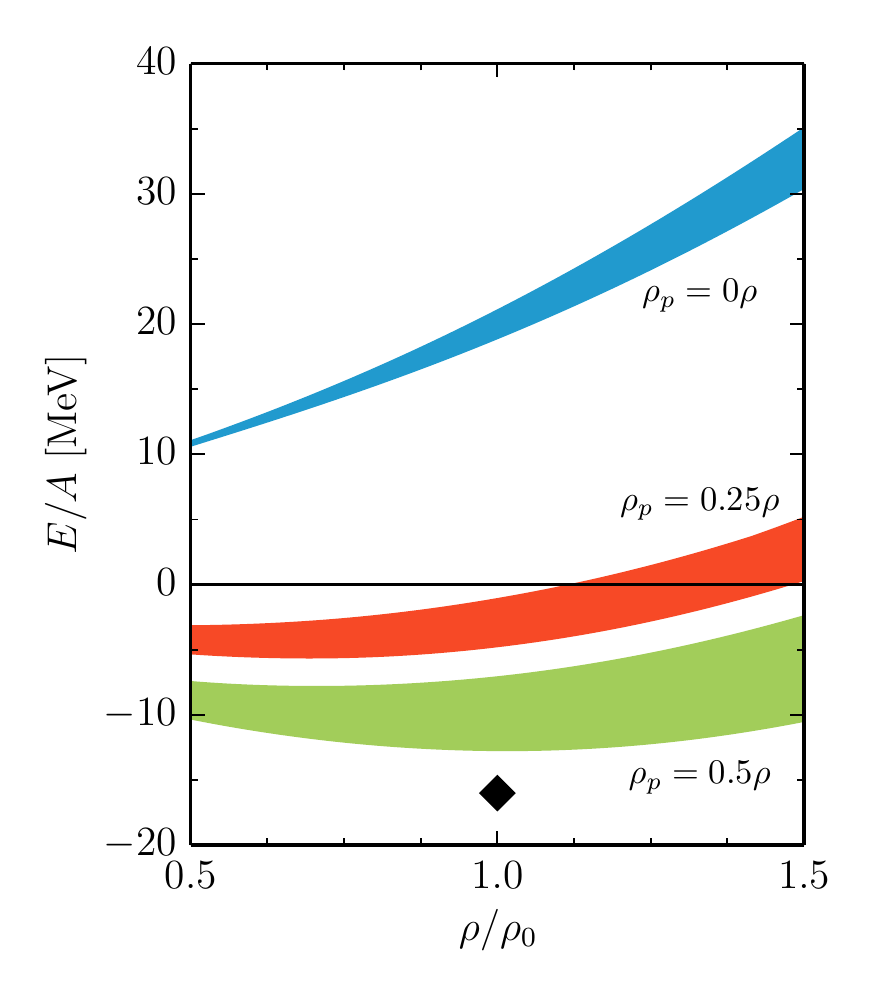}
\caption{Nucleon single-particle potential in symmetric nuclear matter for \(k_F=1.35\ \mathrm{fm}^{-1}\) and \(k_F=1.00\ \mathrm{fm}^{-1}\) (left) and energy per particle of nuclear matter with different proton fractions \(\rho_p/\rho\) (right). 
The bands represent the variation of our results with the cutoff, see text. 
The diamond symbolizes the empirical saturation point of symmetric nuclear matter.}
\label{fig:N}
\end{figure*}

\begin{table*}[htpb]
\caption{\(\Lambda\) single-particle potential \(U_\Lambda(k=0)\) in symmetric nuclear matter at saturation density, \(k_F=1.35\ \mathrm{fm}^{-1}\). Values are given in MeV and decomposed into partial wave contributions.}
\label{tab:U0L}
\centering
\begin{tabular}{c>{$}c<{$}>{$}c<{$}>{$}c<{$}>{$}c<{$}>{$}c<{$}>{$}c<{$}>{$}c<{$}}
\toprule
$U_\Lambda(k=0)$ & {}^1S_0 & {}^3S_1$+${}^3D_1 & {}^3P_0 & {}^1P_1 & {}^3P_1 & {}^3P_2$+${}^3F_2 & \text{Total} \\
\cmidrule(lr){1-1} \cmidrule(lr){2-7} \cmidrule(lr){8-8}
NLO (500) cont & -15.4 & -15.7 & 1.0 & 1.8 & 1.5 & -1.3 & -28.3 \\
NLO (550) cont & -13.9 & -12.7 & 0.9 & 1.6 & 1.5 & -1.2 & -24.2 \\
NLO (600) cont & -12.9 & -13.5 & 0.8 & 1.3 & 1.4 & -1.2 & -24.4 \\
NLO (650) cont & -12.4 & -16.3 & 0.7 & 1.2 & 1.3 & -1.2 & -27.0 \\
\cmidrule(lr){1-1} \cmidrule(lr){2-7} \cmidrule(lr){8-8}
LO (600) gap & -12.1 & -25.9 & -1.7 & 1.5 & 1.7 & -0.4 & -37.2 \\
LO (600) cont & -13.2 & -28.0 & -1.9 & 1.5 & 1.7 & -0.4 & -40.7 \\
NLO (600) gap & -13.1 & -13.9 & 0.9 & 1.3 & 1.4 & -1.2 & -24.8 \\
NSC97f gap & -14.7 & -24.1 & 0.4 & 2.4 & 4.1 & -0.8 & -34.1 \\
NSC97f cont & -14.5 & -25.2 & 0.4 & 2.3 & 3.9 & -0.9 & -35.5 \\
J\"ulich '04 gap & -10.5 & -36.5 & -0.7 & -0.6 & 0.5 & -3.2 & -51.7 \\
J\"ulich '04 cont & -11.2 & -38.0 & -0.7 & -0.7 & 0.5 & -3.3 & -54.2 \\
\bottomrule
\end{tabular}
\end{table*}

\begin{figure*}[htpb]
\centering
\includegraphics[scale=\figscale]{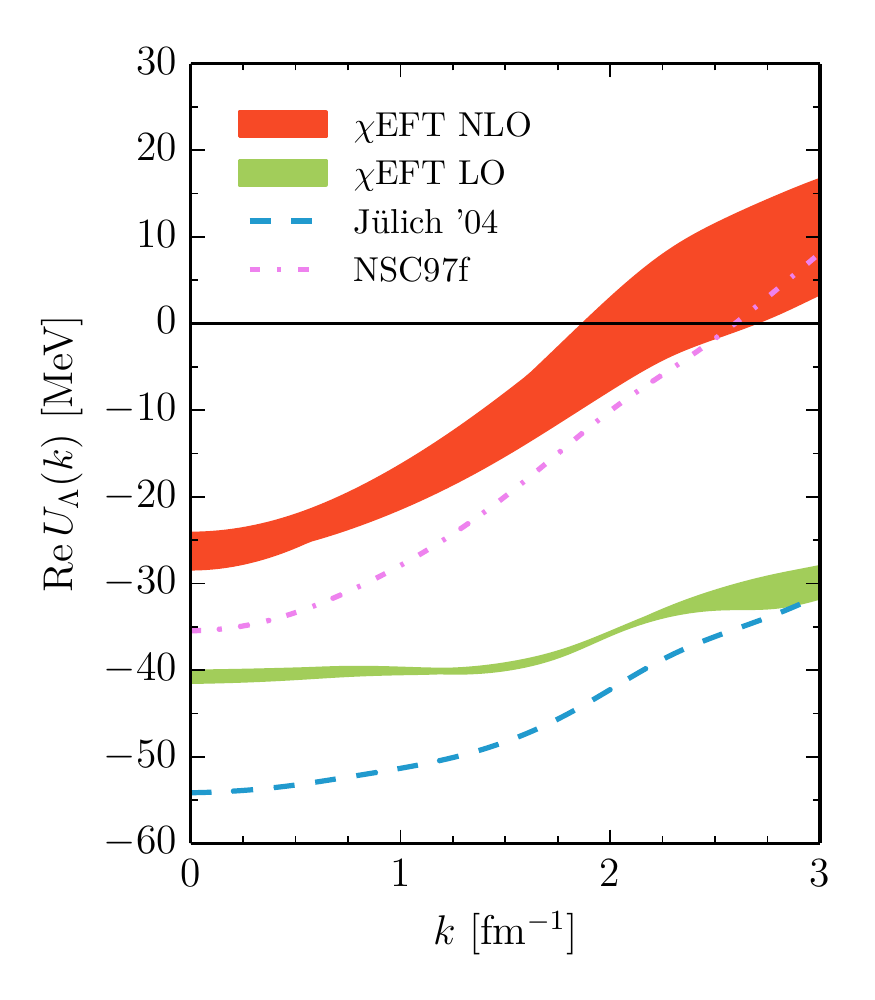} \qquad \qquad
\includegraphics[scale=\figscale]{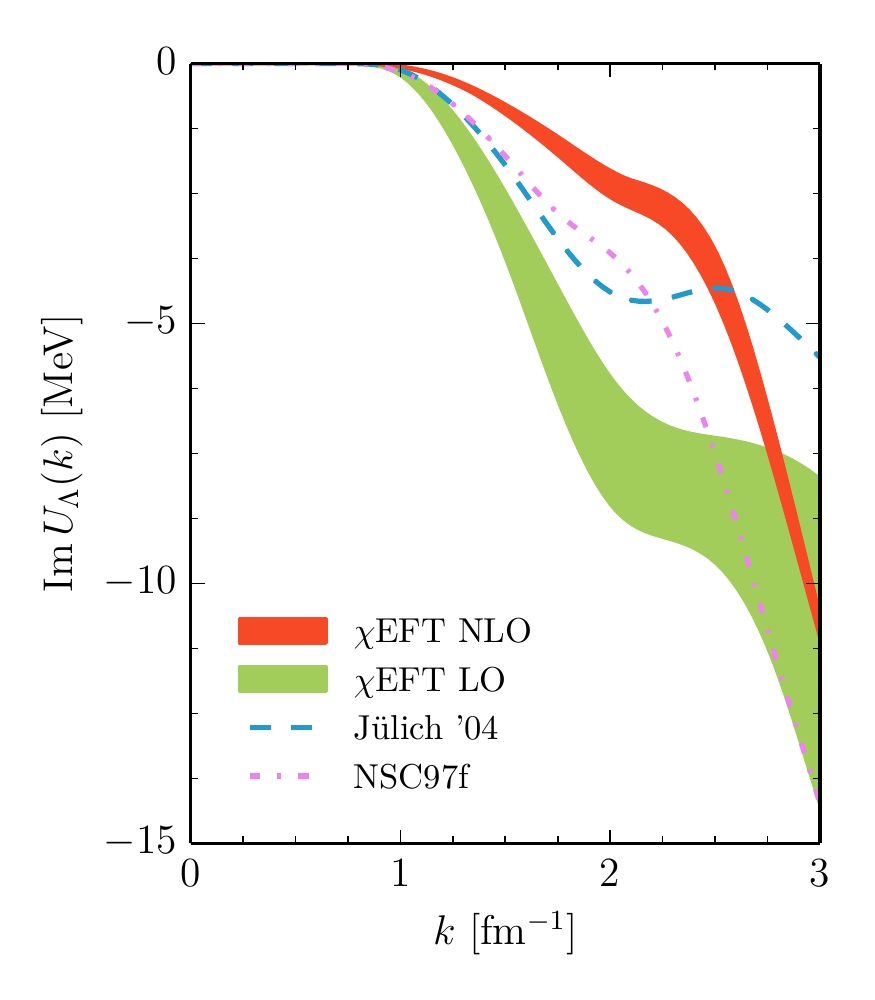}
\caption{Momentum dependence of the real and imaginary parts of the single-particle potential of a \(\Lambda\) hyperon in isospin-symmetric nuclear matter at saturation density.
The bands represent the variation of our results with the cutoff, see text. 
}
\label{fig:UL_band}
\end{figure*}

First, we review the results for the nucleon single-particle potential derived from chiral effective field theory at NLO, 
as this is an input for calculations of hyperons in nuclear matter.
Figure~\ref{fig:N} shows \(U_N(k)\) for symmetric nuclear matter at the Fermi momenta \(k_F = 1.35\ \mathrm{fm}^{-1}\) and \(k_F = 1.0\ \mathrm{fm}^{-1}\), corresponding to densities \(\rho=0.166\ \mathrm{fm}^{-3}\) and \(\rho=0.068\ \mathrm{fm}^{-3}\), as determined from the chiral \(NN\) potential and from the Nijmegen 93 model \cite{Stoks1994}.
According to the Hugenholtz--van-Hove theorem the value at \(k=k_F\) has to be \(U_N(k_F)\approx-53\) MeV at saturation density. 
The results of our calculation with the EFT interaction are consistent with this constraint.
Furthermore, fig.~\ref{fig:N} shows the total binding energy per particle % \(E/A = \epsilon / \rho\)
\begin{equation}
\frac EA = \frac{\epsilon}{\rho}\,.
\end{equation}
The particle density \(\rho\) is given by a sum over the baryonic species that occupy Fermi seas,
\begin{equation}
\rho = \sum_B\frac{k^{(B)\,3}_F}{3\pi^2} \,,
\end{equation}
and the energy density \(\epsilon\) can be calculated from the single-particle potential as
\begin{equation}
\epsilon = \sum_B\left(
%\frac{k^{(B)\,3}_F}{3\pi^2} M_B +
\frac{k^{(B)\,5}_F}{10\pi^2M_B} +
\frac1{2\pi^2} \int_0^{k^{(B)}_F} \!\mathrm dk\, k^2\, \Re U_B(k)
\right)\,.
\end{equation}
As it is typical for non-relativistic \(G\)-matrix calculations with % chiral low-momentum 
realistic interactions the empirical saturation point of isospin-symmetric nuclear matter is not reproduced without the inclusion of three-nucleon forces \cite{Jeukenne1976,Li2006}. 
Note that the employed nucleon-nucleon interaction at NLO in chiral EFT becomes too repulsive for higher energies (cf.\ fig.~\ref{fig:Nphase}).
This feature appears to be reflected in the curve for the binding energy per nucleon which saturates at lower densities 
than usually found in calculations using (chiral and other) nucleon-nucleon potentials \cite{Li2006,Bogner2005,Krewald2012,Machleidt2010,Carbone2013}.
But, as expected, our results still lie within the well-known Coester band \cite{Coester1970}. 
In this context we want to stress that we show the nucleonic results only for illustrative purposes. 
Considering the recent arrival of \(NN\) interactions at fifth order in chiral EFT \cite{Epelbaum2014,Entem2015} 
the NLO potential employed here is obviously not state-of-the-art. However, for consistency reasons
we prefer to use \(NN\) and \(YN\) interactions at the same order of the chiral expansion. In any
case, as we will see below, the properties of hyperons in nuclear matter do not depend strongly on the 
nucleon single-particle potential \(U_N(k)\), and therefore the \(NN\) interaction up to NLO is certainly 
sufficient for our purposes.

\begin{table*}[htpb]
\caption{\(\Sigma\) single-particle potential \(U_\Sigma(k=0)\) in symmetric nuclear matter at saturation density, 
\(k_F=1.35\ \mathrm{fm}^{-1}\). Values are given in MeV and decomposed into partial wave contributions.}
\label{tab:U0S}
\centering
\begin{tabular}{c>{$}c<{$}>{$}c<{$}>{$}c<{$}>{$}c<{$}>{$}c<{$}>{$}c<{$}>{$}c<{$}}
\toprule
$U_\Sigma(k=0)$ & {}^1S_0 & {}^3S_1$+${}^3D_1 & {}^3P_0 & {}^1P_1 & {}^3P_1 & {}^3P_2$+${}^3F_2 & \text{Total} \\
\cmidrule(lr){1-1} \cmidrule(lr){2-7} \cmidrule(lr){8-8}
NLO (500) cont & -4.6 & 13.5 & 1.5 & 0.6 & 0.4 & 0.3 & 11.6 \\
NLO (550) cont & -4.2 & 17.3 & 1.3 & 0.4 & 0.2 & -0.0 & 14.9 \\
NLO (600) cont & -4.7 & 15.4 & 1.2 & 0.2 & 0.0 & -0.4 & 11.5 \\
NLO (650) cont & -4.9 & 11.9 & 1.0 & 0.0 & -0.1 & -0.8 & 7.0 \\
\cmidrule(lr){1-1} \cmidrule(lr){2-7} \cmidrule(lr){8-8}
LO (600) gap & -1.8 & 25.3 & -1.9 & -0.2 & -1.4 & -1.1 & 18.7 \\
LO (600) cont & -2.2 & 22.1 & -1.9 & -0.2 & -1.2 & -1.0 & 15.5 \\
NLO (600) gap & -5.6 & 15.4 & 1.1 & 0.1 & -0.2 & -0.6 & 9.9 \\
NSC97f gap & 1.9 & -17.3 & 0.4 & -2.1 & 1.1 & -2.4 & -19.1 \\
NSC97f cont & -0.5 & -17.2 & 0.4 & -2.1 & 1.0 & -2.6 & -22.0 \\
J\"ulich '04 gap & -8.4 & -4.0 & 0.4 & -2.0 & -1.8 & -3.7 & -20.1 \\
J\"ulich '04 cont & -7.9 & -5.6 & 0.4 & -2.0 & -2.5 & -3.8 & -21.9 \\
\bottomrule
\end{tabular}
\end{table*}

\begin{figure*}[htpb]
\centering
\includegraphics[scale=\figscale]{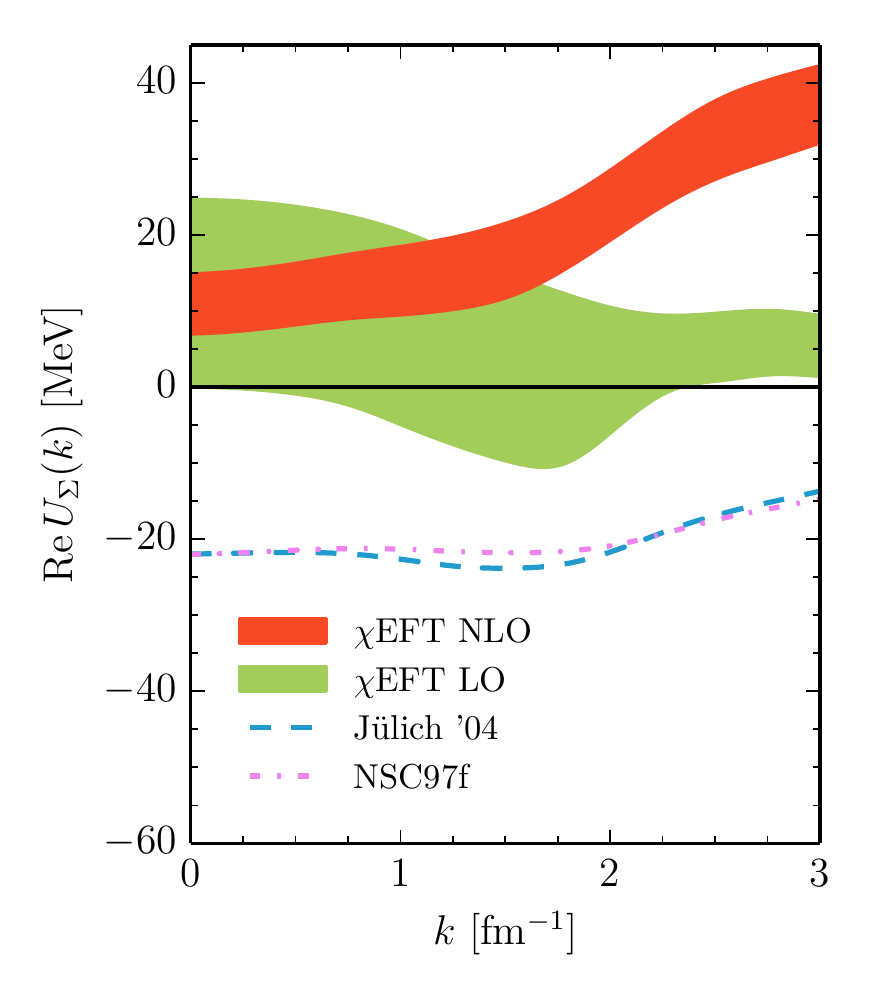} \qquad \qquad
\includegraphics[scale=\figscale]{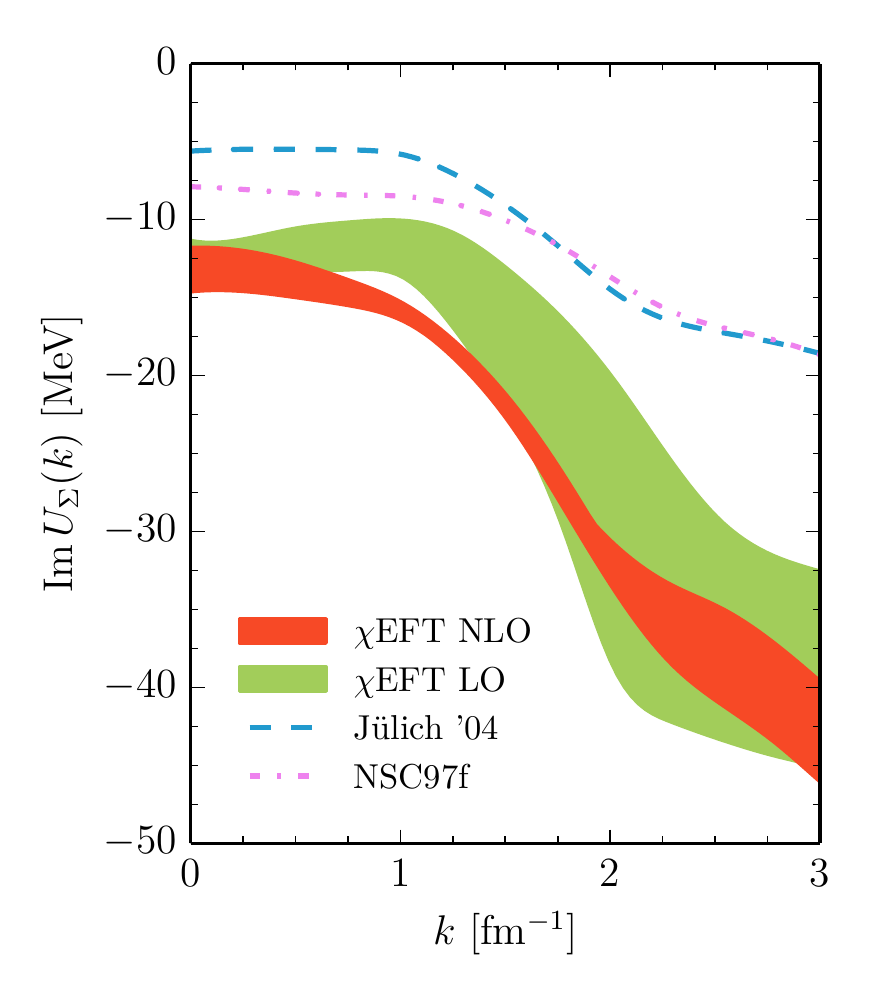}
\caption{Momentum dependence of the real and imaginary parts of the single-particle potential of a \(\Sigma\) hyperon in isospin-symmetric nuclear matter at saturation density.
The bands represent the variation of our results with the cutoff, see text. 
}
\label{fig:US_band}
\end{figure*}

Now we turn to the properties of hyperons in symmetric nuclear matter as they follow from SU(3) chiral effective field theory.
In table~\ref{tab:U0L} values for the depth of the \(\Lambda\) single-particle potential \(U_\Lambda(k=0)\) at saturation density are given.
The results of the LO and NLO results are consistent with the empirical value of about \(-28~\mathrm{MeV}\) as deduced from 
binding energies of \(\Lambda\) hypernuclei \cite{Millener1988,Yamamoto1988}.
The results for the gap choice of intermediate spectra are similar to ref.~\cite{Haidenbauer2015a}, where a phenomenological 
parametrization of the nucleon single-particle potential \(U_N(k)\) has been used.
This suggests that the results for the \(\Lambda\) single-particle potential do not depend strongly on those
of the nucleon in nuclear matter.
(Actually, using this phenomenological parametrization, we can reproduce the
results of ref.~\cite{Haidenbauer2015a} which served as a test for the new code developed for the
present investigation.) 
The differences for \(U_\Lambda(k=0)\) between the gap choice and the continuous choice are a few MeV, comparable to what 
has been found, e.g., in ref.~\cite{Rijken1998} for the Nijmegen NSC97 potentials.
Obviously, the two phenomenological models (J\"ulich~'04, Nijmegen NSC97f) predict more attractive 
values of \(U_\Lambda(0)=(-35\ldots-50)\ \mathrm{MeV}\), 
where the main difference is due to the contribution in the $^3S_1$ partial wave.  
As already discussed in ref.~\cite{Haidenbauer2015a}, 
we believe that this due to the fact that the $\Lambda p$ $^3S_1$$\leftrightarrow$$^3D_1$ transition 
is significantly larger in the NLO chiral EFT interaction as compared to the one of the LO interaction and 
of the phenomenological models whereas the diagonal ($^3S_1$$\leftrightarrow$$^3S_1$ and $^3D_1$$\leftrightarrow$$^3D_1$)
transitions are accordingly smaller. 

The momentum dependence of the real and imaginary parts of the \(\Lambda\) single-particle potential is presented in fig.~\ref{fig:UL_band}.
A marked difference between LO and NLO is, that the \(\Lambda\) single-particle potential at NLO turns to repulsion at fairly
low momenta around \(k\approx 2\ \mathrm{fm}^{-1}\). A similar behavior is also found for the NSC97f potential. 
The cutoff dependence at LO seems to be accidentally weak.

Corresponding results for \(\Sigma\) hyperons in isospin-sym\-metric nuclear matter at saturation density are given 
in table~\ref{tab:U0S} and are also graphically displayed in fig.~\ref{fig:US_band}.
The presented results are for the neutral \(\Sigma^0\) hyperon. The small difference to the results for charged \(\Sigma^+\) and \(\Sigma^-\) hyperons comes solely from the mass difference of the three \(\Sigma\) particles, and is of the order of \((0.5\ldots1)\ \mathrm{MeV}\), where the difference between \(\Sigma^0\) and \(\Sigma^\pm\) is larger than the one between \(\Sigma^+\) and \(\Sigma^-\).
According to analyses of data on \((\pi^-,K^+)\) spectra related to \(\Sigma^-\) formation in heavy nuclei the \(\Sigma\)-nuclear potential
is moderately repulsive in symmetric nuclear matter, see the review~\cite{Friedman2007}.
This feature is well reproduced in our calculation for NLO and even for LO.
Indeed, in the course of constructing the NLO interaction it turned out
that the available $YN$ scattering data could be fitted equally well with an attractive
or a repulsive interaction in the $^3S_1$ partial wave of the $I=3/2$ $\Sigma N$ channel
\cite{Haidenbauer2013a}, which is the partial wave that provides the dominant contribution 
to the \(\Sigma\) single-particle potential, cf. table~\ref{tab:U0S} and also 
table~4 in \cite{Haidenbauer2015a}. For the reasons discussed above, the repulsive solution was adopted.
Note that models derived within the meson-exchange framework often fail to produce a repulsive $\Sigma$-nuclear 
potential and the two phenomenological \(YN\) potentials considered here are exemplary for this deficiency. 
As visible in fig.~\ref{fig:US_band} the \(\Sigma\) potential stays repulsive for higher momenta.
The imaginary part of the \(\Sigma\)-nuclear potential at saturation density is in good agreement with the empirical value of \(-16\ \mathrm{MeV}\) as extracted from \(\Sigma^-\)-atom data \cite{Dover1989}.
The imaginary potential is mainly induced by the \(\Sigma N\) to \(\Lambda N\) conversion in nuclear matter.
Evidently, the bands representing the cutoff dependence of the chiral potentials, become smaller when going to higher order in the chiral expansion.
This feature has been also observed for the \(YN\) scattering observables \cite{Haidenbauer2013a}.

\begin{figure*}[htpb]
\centering
\includegraphics[scale=\figscale]{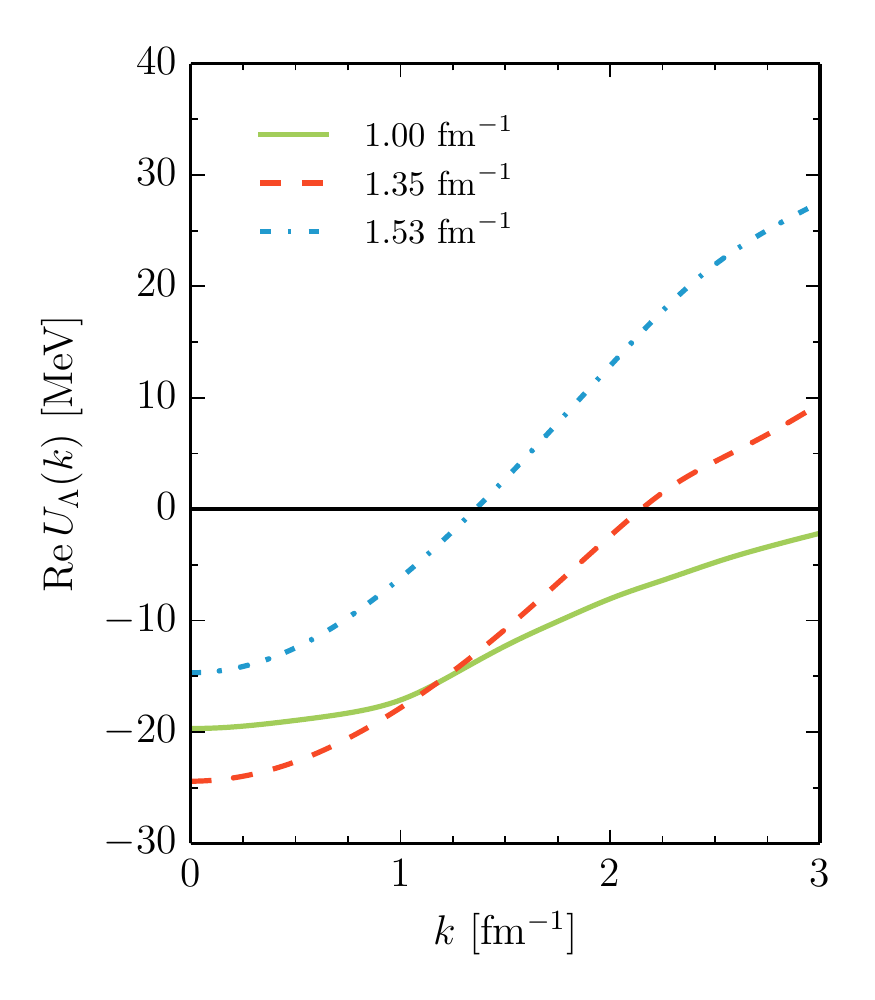} \qquad \qquad
\includegraphics[scale=\figscale]{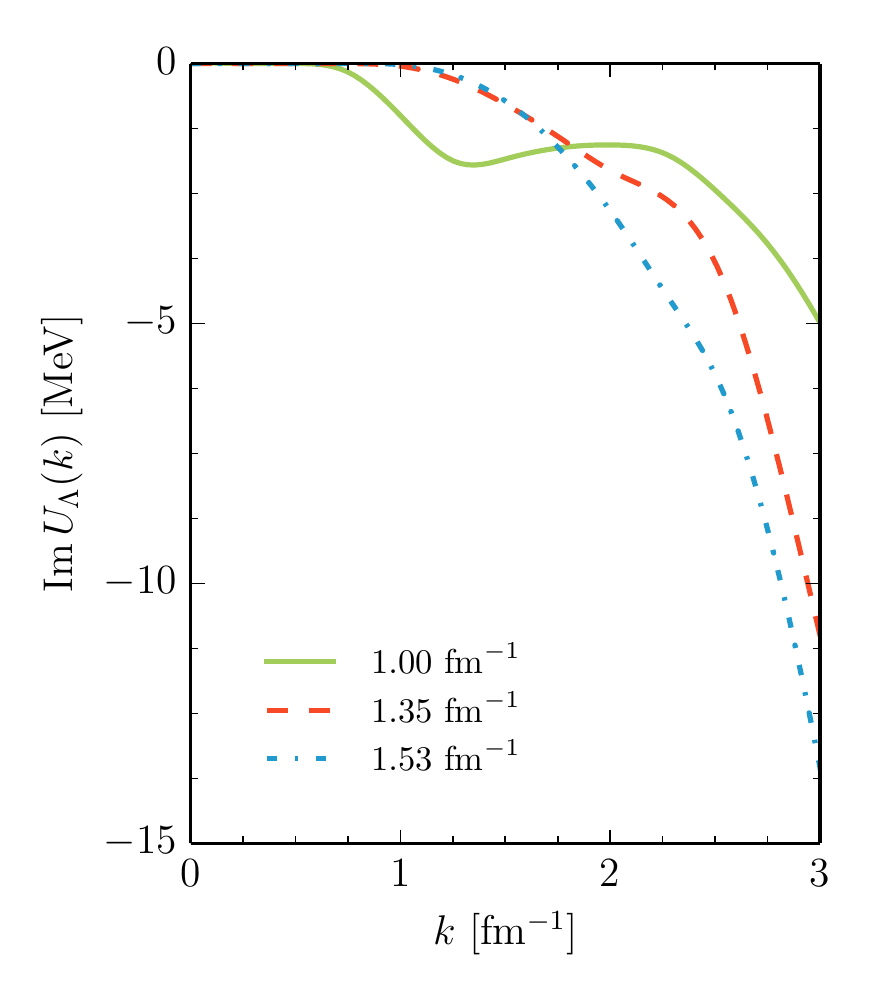}
\caption{Momentum dependence of the real and imaginary parts of the \(\Lambda\) single-particle potential
for different Fermi momenta \(k_F=(1.00,1.35,1.53)\, \mathrm{fm}^{-1}\) in symmetric nuclear matter, 
calculated in \(\chi\)EFT at NLO with a cutoff $\Lambda=600$ MeV.}
\label{fig:UL}
\end{figure*}

\begin{figure*}[htpb]
\centering
\includegraphics[scale=\figscale]{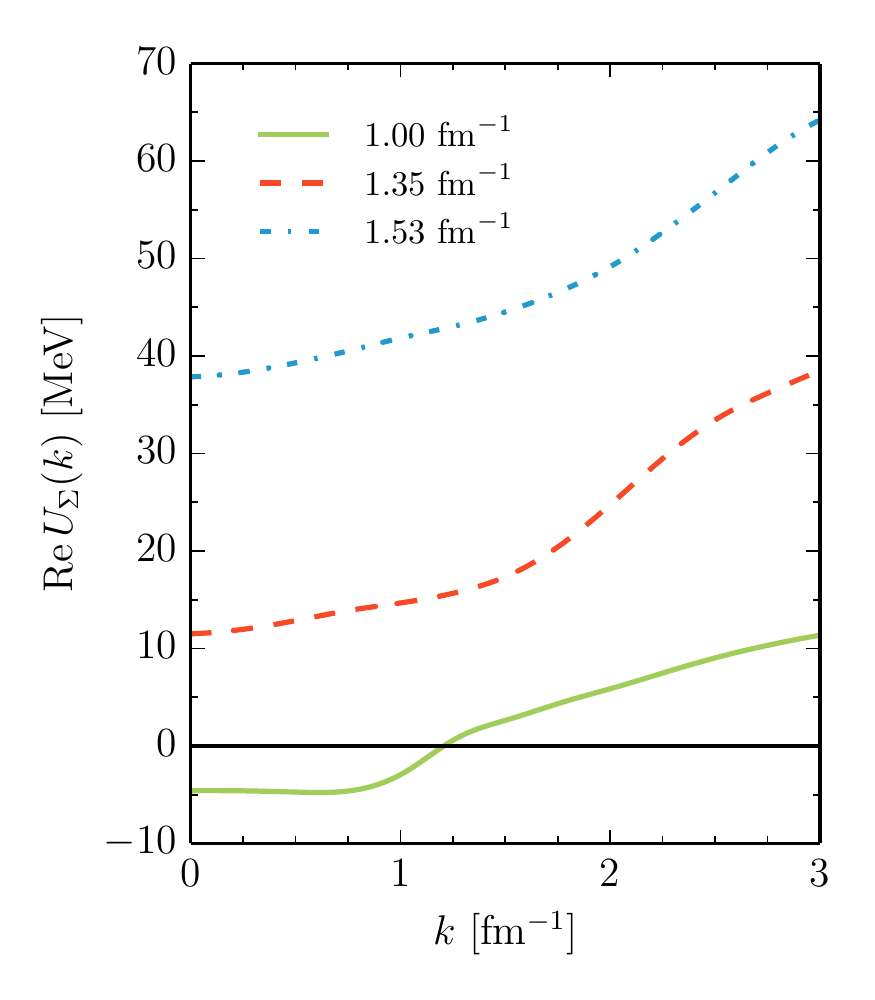} \qquad \qquad
\includegraphics[scale=\figscale]{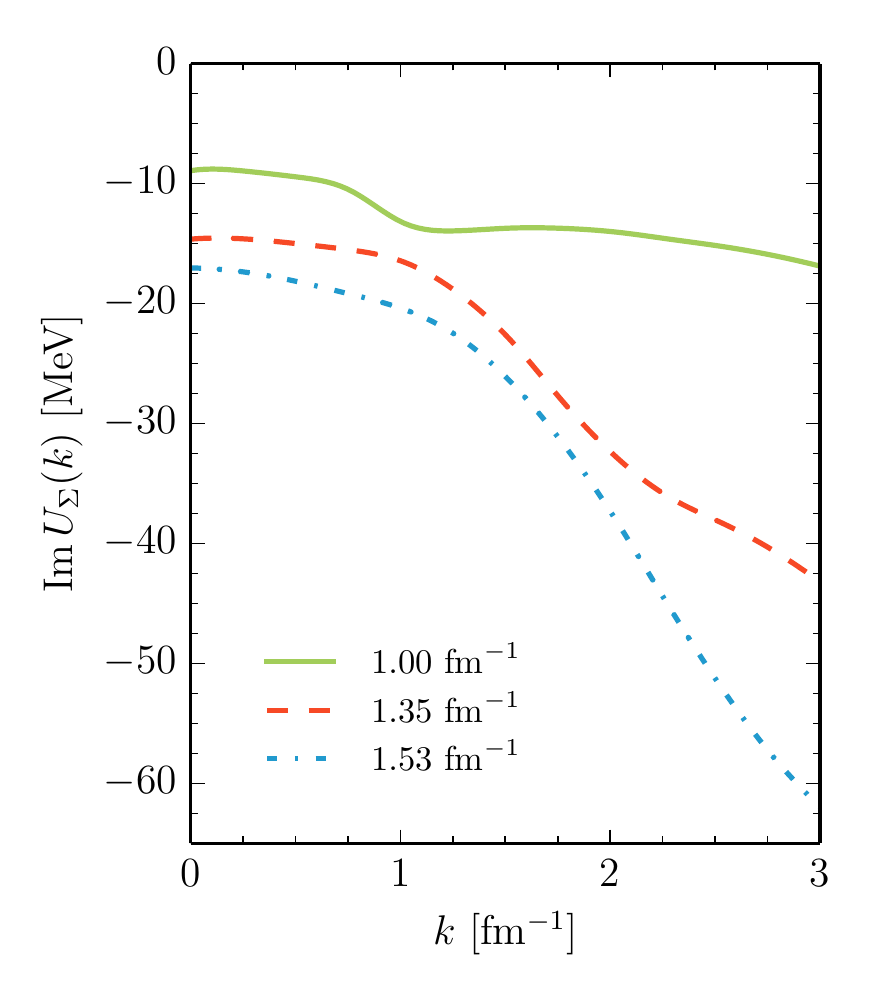}
\caption{Momentum dependence of the real and imaginary parts of the \(\Sigma\) single-particle potential 
for different Fermi momenta \(k_F=(1.00,1.35,1.53)\, \mathrm{fm}^{-1}\) in symmetric nuclear matter, 
calculated in \(\chi\)EFT at NLO with a cutoff $\Lambda=600$ MeV.}
\label{fig:US}
\end{figure*}

\begin{figure*}[htpb]
\centering
\includegraphics[width=\textwidth]{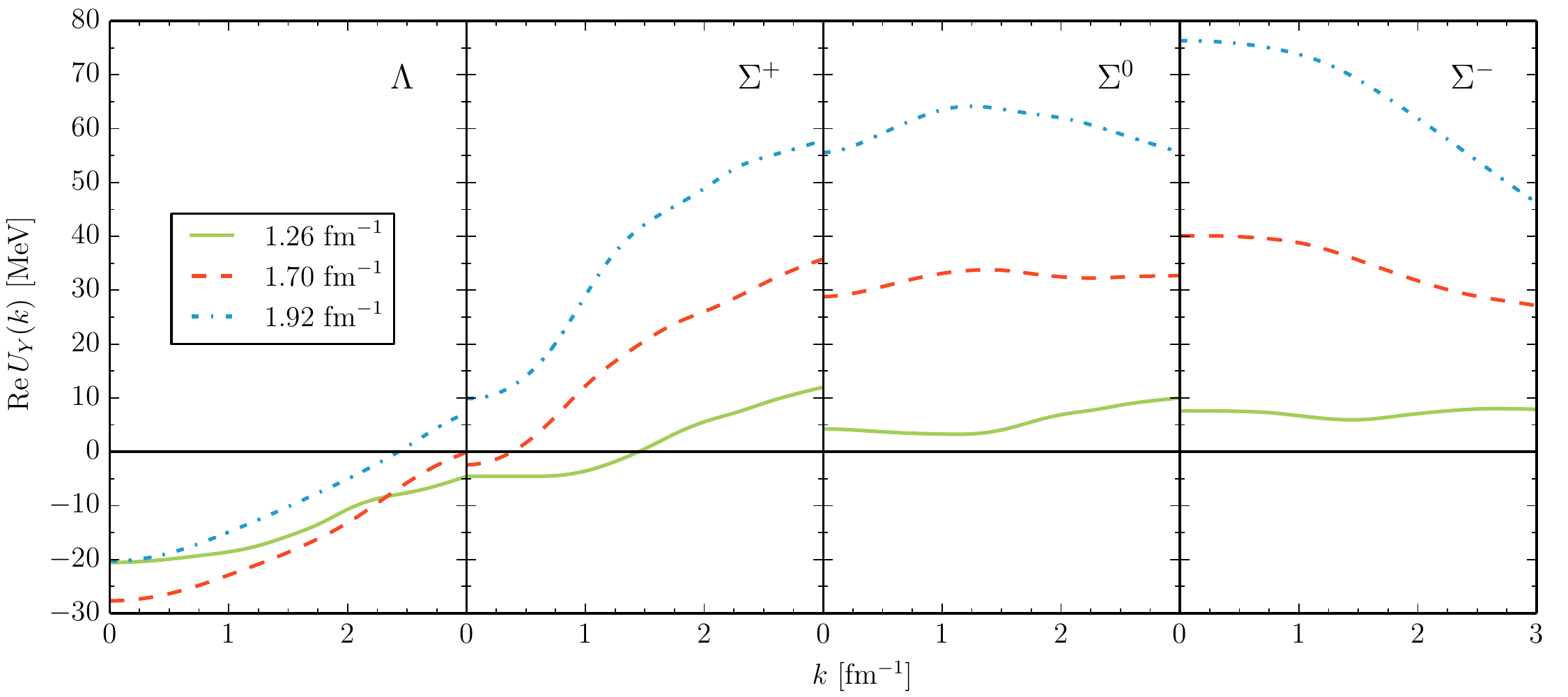}
\caption{Momentum dependence of the real part of the single-particle potentials for hyperons in pure neutron matter for 
different Fermi momenta \(k_F=(1.26,1.70,1.92)\, \mathrm{fm}^{-1}\), 
calculated in \(\chi\)EFT at NLO with a cutoff $\Lambda=600$ MeV.}
\label{fig:Unm}
\end{figure*}

In the following we provide a more detailed view on the dependence of our in-medium results on the densities of protons and neutrons.
The corresponding predictions are shown only for the chiral EFT interaction at NLO 
with a fixed cutoff, namely \(\Lambda= 600\ \mathrm{MeV}\), for reasons of clearer presentation. 
However, one should keep in mind that these results are likewise subject to variations with the cutoff and, 
specifically, in case of a weak sensitivity to the density the latter effect might be actually smaller than 
the cutoff dependence. 

In figs.~\ref{fig:UL} and \ref{fig:US} the dependence of the single-particle 
potential \(U_Y(k)\) for \(\Lambda\) and \(\Sigma\) on the density in symmetric nuclear matter is shown.
The chosen nucleon Fermi momenta \(k_F=(1.00,1.35,1.53)\, \mathrm{fm}^{-1}\) correspond to densities 
of about \(\rho = (0.4,1.0,1.5) \rho_0\) with \(\rho_0 = 0.16\ \mathrm{fm}^{-3}\).
The momentum dependence of the potentials is similar for different densities, but their magnitude 
varies strongly. Especially for the \(\Sigma\) hyperon the single-particle potential can even become attractive at low densities.

Pure neutron matter is another interesting environment for the in-medium behavior of hyperons. Therefore, we display in fig.~\ref{fig:Unm} 
also the density dependence of \(U_Y(k)\) for \(\Lambda\) and \(\Sigma\) hyperons in pure neutron matter.
The Fermi momenta of the neutrons \(k_F=(1.26,1.70,1.92)\, \mathrm{fm}^{-1}\) correspond the same densities 
as selected in figs.~\ref{fig:UL} and \ref{fig:US} for isospin-symmetric nuclear matter. Due to the maximal asymmetry 
between protons and neutrons, the single-particle potentials for the three (\(\Sigma^+,\Sigma^0,\Sigma^-\)) hyperons are rather different.
The \(\Lambda\) single-particle potential \(U_\Lambda(k)\) is slightly more attractive than the one in isospin-sym\-metric nuclear matter at the same density, cf. fig.~\ref{fig:UL}, 
thus indicating only a weak dependence on the composition of nuclear matter.
The small surplus of attraction can be understood from the reduction of Pauli blocking effects in pure neutron matter.

In order to get a more detailed insight into the interaction of hyperons with heavy nuclei, we consider also the strength of the \(\Lambda\)-nuclear 
spin-orbit coupling.
It is experimentally well established \cite{Ajimura2001a,Akikawa2002} that the \(\Lambda\)-nucleus spin-orbit force is very small.
In the following we will present results for the so-called Scheerbaum factor \(S_B\).
It quantifies the strength of the \(\Lambda\)- or \(\Sigma\)-nuclear spin-orbit potential, which takes the form \cite{Scheerbaum1976}
\begin{equation}
U_B^{ls}(r) = -\frac\pi2 S_B\frac1r \frac{\mathrm d\rho(r)}{\mathrm dr}\vec l\cdot\vec\sigma\,,
\end{equation}
where \(\rho(r)\) is the nucleon density distribution, \(\vec l\) the (single-particle) orbital angular momentum operator and \(\vec \sigma\) the hyperon spin operator.
In isospin-symmetric nuclear matter with Fermi momentum \(k_F\) the Scheerbaum factor \(S_{B_1}\) is obtained from the \(G\)-matrix 
elements via the relation~\cite{Fujiwara1999}
\begin{align*}
S_{B_1}(k_{1}) = -\frac{3\pi}{4k_F^3} (1 &+\delta_{B_1B_2}(-1)^{L+S}) \\
 \times \sum_{B_2=n,p} \sum_{J}(2J+ &1) \xi_{12}(1+\xi_{12})^2 \displaybreak[0] \\
 \times \int_{k_\mathrm{min}}^{k_\mathrm{max}}\! \frac{\mathrm dk}{(2\pi)^3}\, W & (k_{1},k)
\operatorname{Re}\Big\{ \\
(J+2) & \, G_{(B_1B_2)(B_1B_2)}^{1J+1,1J+1,J}(k,k;\bar K,\omega_\mathrm{o.s.}) \\
+ & \, G_{(B_1B_2)(B_1B_2)}^{1J,1J,J}(k,k;\bar K,\omega_\mathrm{o.s.}) \\
-(J-1) & \, G_{(B_1B_2)(B_1B_2)}^{1J-1,1J-1,J}(k,k;\bar K,\omega_\mathrm{o.s.}) \\
-\sqrt{J(J+1)} & \, G_{(B_1B_2)(B_1B_2)}^{1J,0J,J}(k,k;\bar K,\omega_\mathrm{o.s.}) \\
-\sqrt{J(J+1)} & \, G_{(B_1B_2)(B_1B_2)}^{0J,1J,J}(k,k;\bar K,\omega_\mathrm{o.s.})
\Big\} \, ,
\numberthis
\label{eq:S}
\end{align*}
with \(k_1\) set to zero in the end.
In table~\ref{tab:SL} we present the Scheerbaum factor \(S_\Lambda\) for the \(\Lambda\) hyperon in symmetric nuclear matter at saturation density.
The values in table~\ref{tab:SL} are in agreement with the earlier results of ref.~\cite{Haidenbauer2015a}. 
The difference between the gap and the continuous choice for intermediate spectra is small. 
This is expected because the Scheerbaum factor involves only contributions from $P$-waves and higher partial waves and it
is known that these are much less sensitive to the treatment of the intermediate spectra than the $S$-waves \cite{Rijken1998}.
We use the same \(YN\) interaction as in ref.~\cite{Haidenbauer2015a} where the strength of the antisymmetric spin-orbit contact interaction, 
generating a spin singlet-triplet mixing (\({}^1P_1\leftrightarrow {}^3P_1\)), has been tuned to achieve 
\(S_\Lambda\approx -3.7\ \mathrm{MeV}\, \mathrm{fm}^{5}\), in accordance 
with estimates for the empirical value that is expected to be in the range of around \(-4.6\) to \(-3.0\) \(\mathrm{MeV}\, \mathrm{fm}^{5}\)
\cite{Kohno2010,KohnoPriv}.
In table~\ref{tab:SS} we summarize our results for the Scheerbaum factor \(S_\Sigma\), which are close to the values reported in ref.~\cite{Haidenbauer2015a}.
As before, results are only given for \(\Sigma^0\). The difference among the Scheerbaum factors for the three \(\Sigma\) hyperons 
due to their mass splitting is smaller than 0.5 \(\textrm{MeV}\, \textrm{fm}^{5}\).
In contrast to the leading-order approximation, at NLO negative values of \(S_\Sigma\) are always obtained, similar to the results found with the NSC97f and J\"ulich~'04 models.

\begin{table*}[htpb]
\caption{Scheerbaum factor \(S_\Lambda(k=0)\) in symmetric nuclear matter at saturation density, \(k_F=1.35\ \mathrm{fm}^{-1}\). Values are given in \(\textrm{MeV}\, \textrm{fm}^{5}\) and decomposed into partial wave contributions.}
\label{tab:SL}
\centering
\begin{tabular}{c>{$}c<{$}>{$}c<{$}>{$}c<{$}>{$}c<{$}>{$}c<{$}>{$}c<{$}>{$}c<{$}>{$}c<{$}}
\toprule
$S_\Lambda(k=0)$ & {}^3P_0 & {}^3D_1 & {}^3P_1 & {}^1P_1\leftrightarrow{}^3P_1 & {}^3P_2 & {}^3D_2 & {}^3D_3 & \text{Total} \\
\cmidrule(lr){1-1} \cmidrule(lr){2-8} \cmidrule(lr){9-9}
NLO (500) cont & -5.6 & -0.6 & -4.4 & 10.4 & -3.5 & 0.4 & 0.2 & -3.0 \\
NLO (550) cont & -4.9 & -0.6 & -4.2 & 9.2 & -3.2 & 0.4 & 0.2 & -3.1 \\
NLO (600) cont & -4.4 & -0.6 & -4.0 & 8.3 & -3.1 & 0.4 & 0.2 & -3.2 \\
NLO (650) cont & -4.0 & -0.6 & -3.8 & 7.3 & -3.1 & 0.4 & 0.2 & -3.6 \\
\cmidrule(lr){1-1} \cmidrule(lr){2-8} \cmidrule(lr){9-9}
LO (600) gap & 9.4 & -0.2 & -4.9 & 0.0 & -1.1 & 0.4 & -0.1 & 3.5 \\
LO (600) cont & 10.2 & -0.2 & -4.7 & 0.0 & -1.2 & 0.4 & -0.1 & 4.5 \\
NLO (600) gap & -4.7 & -0.6 & -4.1 & 8.4 & -3.1 & 0.4 & 0.2 & -3.4 \\
NSC97f gap & -2.2 & 0.5 & -11.4 & 2.0 & -2.6 & 1.1 & -2.0 & -14.2 \\
NSC97f cont & -2.0 & 0.5 & -10.8 & 1.9 & -2.8 & 1.2 & -2.0 & -13.8 \\
J\"ulich '04 gap & 4.0 & 0.4 & -1.4 & 5.1 & -9.1 & 0.6 & -1.0 & -1.3 \\
J\"ulich '04 cont & 4.1 & 0.5 & -1.3 & 5.1 & -9.3 & 0.6 & -1.1 & -1.4 \\
\bottomrule
\end{tabular}
\end{table*}

\begin{table*}[htpb]
\caption{Scheerbaum factor \(S_\Sigma(k=0)\) in symmetric nuclear matter at saturation density, \(k_F=1.35\ \mathrm{fm}^{-1}\). Values are given in \(\textrm{MeV}\, \textrm{fm}^{5}\) and decomposed into partial wave contributions.}
\label{tab:SS}
\centering
\begin{tabular}{c>{$}c<{$}>{$}c<{$}>{$}c<{$}>{$}c<{$}>{$}c<{$}>{$}c<{$}>{$}c<{$}>{$}c<{$}}
\toprule
$S_\Sigma(k=0)$ & {}^3P_0 & {}^3D_1 & {}^3P_1 & {}^1P_1\leftrightarrow{}^3P_1 & {}^3P_2 & {}^3D_2 & {}^3D_3 & \text{Total} \\
\cmidrule(lr){1-1} \cmidrule(lr){2-8} \cmidrule(lr){9-9}
NLO (500) cont & -7.7 & 0.6 & -1.0 & -10.6 & 0.9 & 0.1 & 0.1 & -17.7 \\
NLO (550) cont & -6.6 & 0.3 & -0.6 & -9.7 & -0.0 & 0.1 & 0.1 & -16.5 \\
NLO (600) cont & -5.9 & 0.1 & -0.1 & -9.2 & -1.0 & 0.1 & 0.0 & -16.0 \\
NLO (650) cont & -5.3 & 0.1 & 0.2 & -8.3 & -2.0 & 0.1 & -0.0 & -15.3 \\
\cmidrule(lr){1-1} \cmidrule(lr){2-8} \cmidrule(lr){9-9}
LO (600) gap & 9.5 & 0.3 & 3.2 & 0.0 & -2.8 & 0.1 & -0.4 & 9.8 \\
LO (600) cont & 8.7 & 0.3 & 2.6 & 0.0 & -2.4 & 0.1 & -0.3 & 8.9 \\
NLO (600) gap & -5.6 & 0.1 & 0.4 & -10.1 & -1.3 & 0.1 & -0.1 & -16.4 \\
NSC97f gap & -2.0 & 0.1 & -2.8 & -2.7 & -6.2 & 0.3 & -1.5 & -14.9 \\
NSC97f cont & -1.8 & 0.0 & -2.5 & -2.6 & -7.0 & 0.3 & -1.6 & -15.5 \\
J\"ulich '04 gap & -2.2 & 0.1 & 4.5 & -8.2 & -9.6 & 0.3 & -1.0 & -16.4 \\
J\"ulich '04 cont & -2.2 & 0.1 & 6.0 & -9.8 & -9.9 & 0.3 & -1.0 & -16.9 \\
\bottomrule
\end{tabular}
\end{table*}

\begin{figure}[t]
\centering
\includegraphics[scale=\figscale]{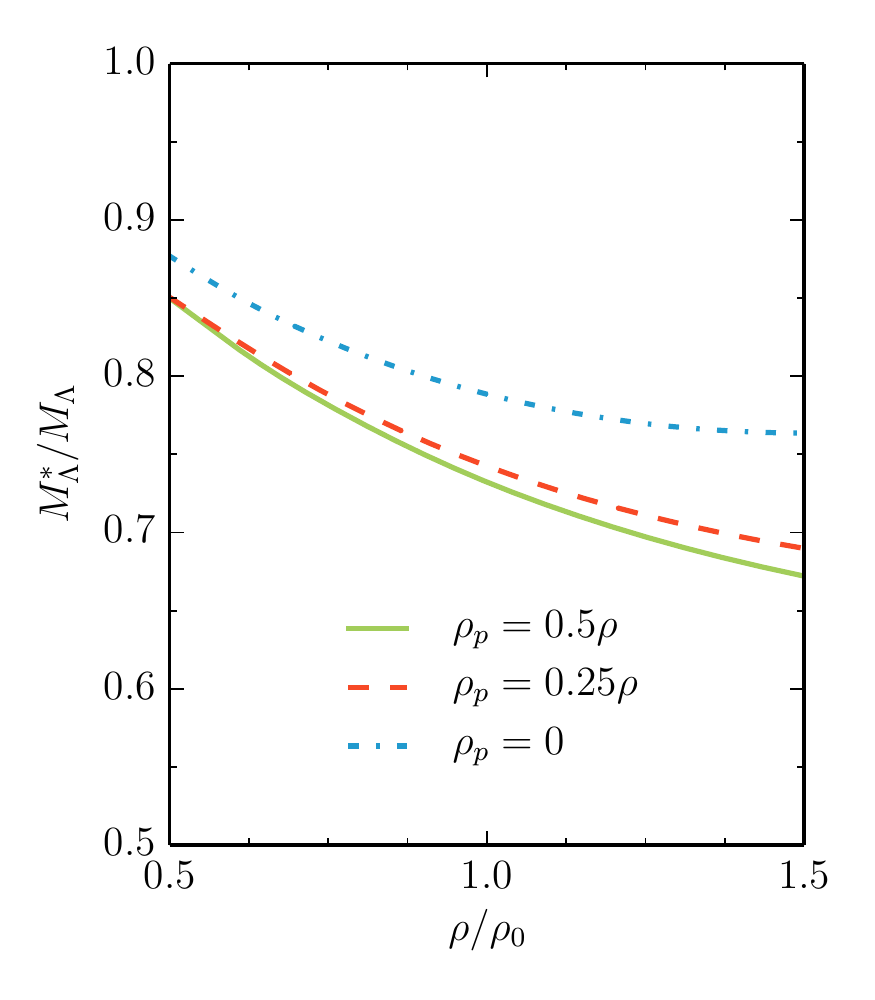}
\caption{Density dependence of the effective mass of a \(\Lambda\) hyperon in (a)symmetric nuclear matter, 
calculated in \(\chi\)EFT at NLO with a cutoff $\Lambda=600$ MeV.}
\label{fig:effm}
\end{figure}

Another important quantity to characterize in-medium properties is the effective baryon mass.
The ratio between the effective and the free hyperon mass in nuclear matter is usually defined as
\begin{equation}
\frac{M^*_B}{M_B} = \left[1+2M_B \left. \frac{ \partial \Re U_B(k) }{ \partial k^2 } \right\vert_{k=0}\right]^{-1} \, .
\end{equation}
We follow ref.~\cite{Kohno1999} where this ratio is computed as
\begin{equation}
\frac{M^*_B}{M_B} = \left[1+\frac{2M_B}{k^2}\Re ( U_B(k)-U_B(0))\right]^{-1} \, , 
\end{equation}
with \(k\approx 1\ \mathrm{fm}^{-1}\). In some works \(k\) is set to the Fermi momentum \(k_F\) in symmetric nuclear matter.
% and/or {fitted to $U$ up to roughly 1 \(\mathrm{fm}^{-1}\)}.
In fig.~\ref{fig:effm} the density dependence of the effective mass \(M^*_\Lambda\) of a \(\Lambda\) hyperon for different isospin asymmetries \(\rho_p = (0,0.25,0.5)\rho\) is shown, where \(\rho_p\) is the proton density and \(\rho\) the total density.
The effective \(\Lambda\) mass in pure neutron matter is slightly higher than the one in symmetric nuclear matter, but the shape of the curves in fig.~\ref{fig:effm} does not depend on the composition of nuclear matter.
Since the momentum-dependence of the \(\Sigma\)-nuclear potentials as obtained from the \(G\)-matrix calculations is not close to a quadratic behavior for momenta \(0\leq k\leq 3\ \mathrm{fm}^{-1}\), the effective mass does not serve as a significant quantity.

\begin{figure*}[htpb]
\centering
\includegraphics[width=\textwidth]{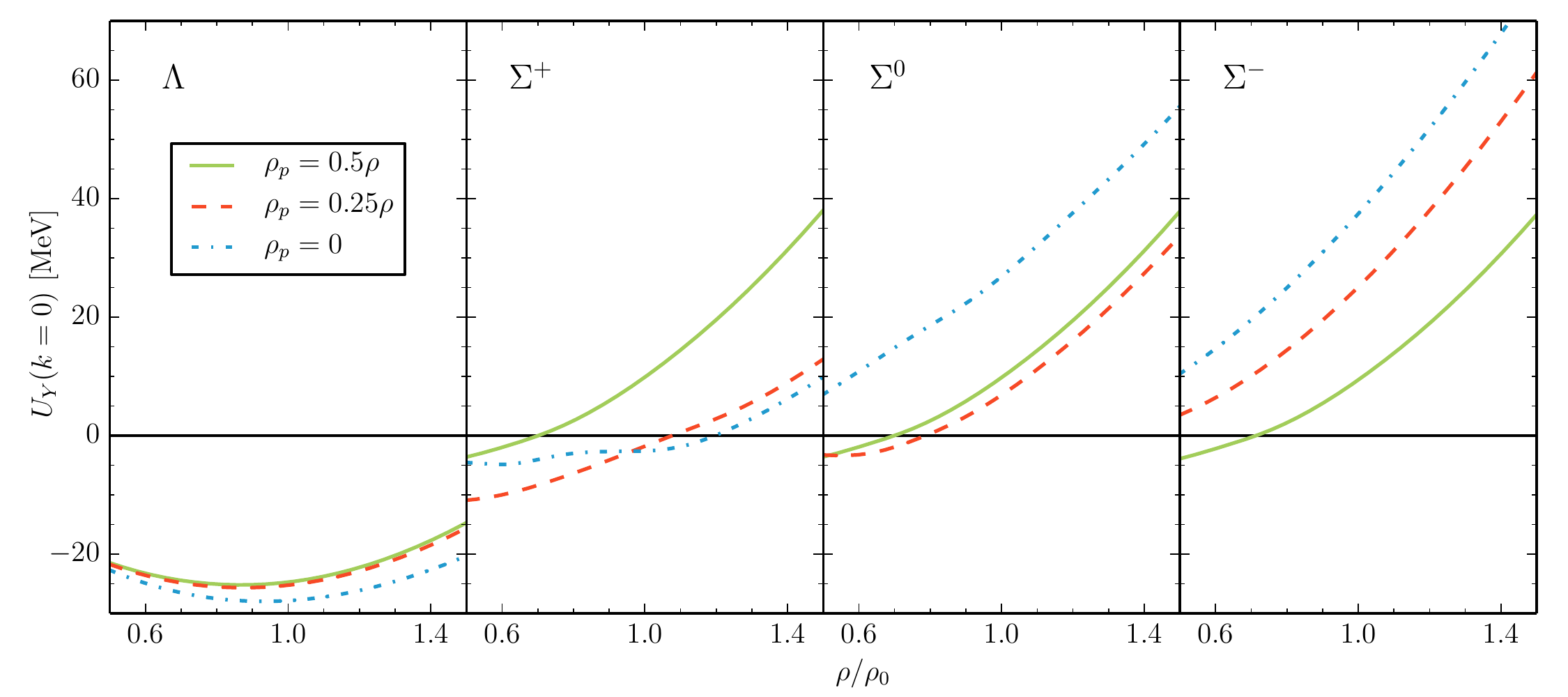}
\caption{Density dependence of the hyperon single-particle potentials at \(k=0\) with different compositions of the nuclear matter, 
calculated in \(\chi\)EFT at NLO with a cutoff $\Lambda=600$ MeV.}
\label{fig:rho}
\end{figure*}

Finally, we investigate in more detail the influence of the composition and density of nuclear matter on the single-particle potentials of hyperons.
In fig.\,\ref{fig:rho} the density dependence of the depth of the nuclear mean-field of \(\Lambda\) or \(\Sigma\) hyperons at rest is shown for isospin-symmetric nuclear matter, asymmetric nuclear matter with \(\rho_p=0.25\rho\) and pure neutron matter.
The single-particle potential of the \(\Lambda\) hyperon is almost independent of the composition of the nuclear medium, because of its isosinglet nature. Furthermore, it is attractive over the whole considered range of density \(0.5\leq \rho/\rho_0\leq 1.5\).
The three \(\Sigma\) hyperons possess (up to small differences from the mass splittings) the same single-particle potential in symmetric nuclear matter. It is attractive for low densities, but turns into repulsion at \(\rho\approx0.7\rho_0\) and stays repulsive for higher densities.
When introducing isospin asymmetry in the nuclear medium a splitting of the single-particle potentials occurs due to the strong isospin dependence of the \(\Sigma N\) interaction.
The splittings among the \(\Sigma^+\), \(\Sigma^0\) and \(\Sigma^-\) potentials as obtained in our microscopic calculation have 
a non-linear dependence on the isospin asymmetry which goes beyond the usual (linear) parametrization in terms of an isovector Lane potential \cite{Dabrowski1999}.

\section{Summary and outlook}
\label{sec:sum}

We have investigated the properties of hyperons in nuclear matter at variable density and isospin asymmetry within the self-consistent Brueckner-Hartree-Fock approach using the continuous choice for intermediate spectra.
The employed microscopic potentials for the hyperon-nucleon interaction are derived within SU(3) chiral effective field theory at next-to-leading order.
In addition the relevant nucleon-nucleon interaction is taken as well from chiral EFT.
We have presented results for the complex-valued single-particle potentials of \(\Lambda\) and \(\Sigma\) hyperons and have studied the density dependence of the hyperon potential depths and the \(\Lambda\) effective mass.
Our results reveal that the hyperon-nucleon interaction in chiral effective field theory is constructed in such way that it is not only consistent with \(YN\) scattering data, but also in agreement with empirical properties of the hyperon-nuclear potentials in symmetric nuclear matter.
In particular, the potential depth \(U_\Lambda(0)\approx-28\ \mathrm{MeV}\) of the \(\Lambda\) hyperon is reproduced and the \(\Sigma\) single-particle potential comes out repulsive in accordance with recent measurements of \(\Sigma^-\) formation spectra on heavy nuclei.
Furthermore, the experimentally indicated very small value of the \(\Lambda\)-nuclear spin orbit coupling is reproduced.

The presented calculations provide a basis for studies of hypernuclei. 
An extension of the calculations to higher densities of about 2-3\(\rho_0\)
will allow for a first exploratory study of neutron star matter 
including hyperons from the viewpoint of chiral effective field theory.
In particular, the inclusion of three-baryon forces, especially the \(\Lambda NN\) interaction, 
represents a very interesting forthcoming extension of these calculations which 
can now be performed consistently within chiral effective field theory since the SU(3) 
classification of the three-baryon forces has been completed \cite{Petschauer2013b,PetschauerPrep}.

\acknowledgement
This work is supported in part by the DFG and the NSFC through funds provided to the Sino-German CRC~110 ``Symmetries and the Emergence of Structure in QCD''.

% \bibliographystyle{h-physrev5-mod} %h-physrev5-mod, phaip, alpha, plain, abbrv
% \bibliography{../../Mendeley/BibTeX/zPaperBib-2015Matter}

%\begin{table}
%\label{tab:1}
%\begin{tabular}{lll}
%\hline\noalign{\smallskip}
%\end{tabular}
%\end{table}

%\begin{thebibliography}{}
%\bibitem{RefJ} % Format for Journal Reference
%\bibitem{RefB} % Format for books
%Author, \textit{Book title} (Publisher, place year) page numbers
%\end{thebibliography}

%The  words  section(s),  equation(s),  figure(s)  and  reference(s)  are  abbreviated  as sect(s)., fig(s)., eq(s).  and ref(s).
%unless they are the first word of a sentence. The word table is always written in full.
%Latin expressions, such as, e.g., i.e., et al., versus (vs.) should be set in italic.

\end{document}